\documentclass[pra,twocolumn,showpacs,floatfix,superscriptaddress,aps]{revtex4}
\usepackage{amsmath}
\usepackage{makeidx}
\usepackage{amssymb}
\usepackage{graphicx}
\usepackage{array}
\usepackage[usenames]{color}
\usepackage[normalem]{ulem}

\usepackage{hyperref}

\begin{document}

\title{Spontaneous symmetry breaking 
in a spin-orbit coupled $f=2$ spinor condensate}

\author{Sandeep Gautam\footnote{sandeepgautam24@gmail.com}}
\author{S. K. Adhikari\footnote{adhikari44@yahoo.com, URL  http://www.ift.unesp.br/users/adhikari}}
\affiliation{Instituto de F\'{\i}sica Te\'orica, Universidade Estadual
             Paulista - UNESP, \\ 01.140-070 S\~ao Paulo, S\~ao Paulo, Brazil}
      

\date{\today}
\begin{abstract}

We study the ground-state density profile of a spin-orbit coupled $f=2$ spinor 
condensate in a quasi-one-dimensional trap. The Hamiltonian of the system is 
invariant under time reversal but not under parity. We identify different 
parity- and time-reversal-symmetry-breaking states. The time-reversal-symmetry 
breaking is possible for degenerate states. A phase separation among densities 
of different components is possible in the domain of time-reversal-symmetry 
breaking. Different types of parity- and time-reversal-symmetry-breaking states 
are predicted analytically and studied numerically. We employ numerical and 
approximate analytic solutions of a mean-field model in this investigation to 
illustrate our findings. 
\end{abstract}

\maketitle

\section{Introduction}
Since the first experimental realization of the spinor Bose-Einstein 
condensate (BEC) with a gas of $^{23}$Na atoms trapped in an optical trap 
\cite{Stamper-Kurn}, a lot of theoretical and experimental studies have been 
done on these systems \cite{ueda,kurn-ueda}. On the theoretical front, 
mean-field theories have been developed both for $f = 1$ \cite{Ohmi,Ho} and 
$f=2$ spinor Bose-Einstein condensates (BECs) \cite{Koashi,Ciobanu}. In our 
previous work \cite{sandeep}, we studied the ground state structure of an $f=1$ 
spin-orbit (SO) coupled spinor BEC with fixed magnetization in a 
quasi-one-dimensional (quasi-1D) trap \cite{luca} within the framework of the 
mean-field theory. In the present work, we extend this study to study an SO 
coupled $f=2$ spinor condensate in a quasi-1D trap. The SO coupling, which 
relies on the generation of the non-Abelian gauge potentials coupling the 
neutral atoms \cite{Osterloh}, can be experimentally realized by controlling 
the atom-light interaction. A variety of SO couplings can be engineered by 
Raman dressing the hyper-fine states. The parameters of atom-light interaction
Hamiltonian, and hence those of coupling can be controlled independently 
\cite{young}. The SO interaction with equal strengths of Rashba \cite{Bychkov} and
Dresselhaus \cite{Dresselhaus,Liu} couplings has been achieved recently 
\cite{Lin,Linx}. The experimentalists employed a pair of Raman lasers to create 
a momentum-sensitive coupling between two internal atomic states of $^{87}$Rb
\cite{Lin,Linx}. This has lead to a flurry of other experiments on SO-coupled 
pseudo-spinor BECs \cite{JY_Zhang}. The generation of SO coupling involving
the three hyper-fine spin components of an $f=1$ spinor condensate using Raman 
dressing has also been studied theoretically \cite{Juzeliunas,lan}. Recently,
SO coupling has also been experimentally realized in degenerate
Fermi gases of $^{40}$K and $^{6}$Li \cite{P_Wang}.

Wang {\em et al.} \cite{Wang} studied theoretically the ground states of 
pseudo-spin-$1/2$ two-component BEC with SO coupling and of three-component 
$f=1$ spinor BEC. In the presence of SO coupling, the ground states of 
all spinor BECs $-$ pseudo-spin-$1/2$ \cite{Gopalakrishnan}, 
$f=1$ \cite{Ruokokoski}, and $2$ \cite{Kawakami} may exhibit different 
types of nontrivial density distribution. In the presence of a uniform magnetic 
field the ground states of  $f = 1$ \cite{Ohmi,Ho,Zhou,Matuszewski} and $f = 2$ 
\cite{Ciobanu,ueda,GP-Zheng} spinor BECs exhibit interesting behavior including 
the possibility of a phase separation \cite{Matuszewski,Lin} among different 
spin components. There have also been different static and dynamic studies in 
SO-coupled BECs, such as, Josephson oscillation \cite{josep}, intrinsic 
spin-Hall effect \cite{Larson}, solitons \cite{sol}, force on a moving 
impurity \cite{He}, chiral confinement \cite{Merkl}, and superfluidity 
\cite{super}, etc.

In this paper, we investigate the ground state of an SO-coupled $f=2$ spinor
BEC in a quasi-1D trap for an arbitrary magnetization. The Hamiltonian of this 
system is invariant under time reversal $\cal T$ but not under parity. 
Consequently, different types of parity-breaking states are found. 
Time-reversal symmetry-breaking states are also found in the presence of 
degeneracy. In the absence of degeneracy, the states preserve time-reversal 
symmetry. The five spin-component wave functions of the $f = 2$ spinor
condensate satisfy a coupled mean-field Gross-Pitaevskii (GP) equation 
with  three interaction parameters: $c_0 \propto (4a_2+3a_4)/7$,
$c_1 \propto (a_4-a_2)/7$, and  $c_2 \propto (7a_0-10a_2+3a_4)/7$, where
$a_0,~a_2$, and $a_4$ are the $s$-wave scattering lengths in total spin 
$f_{\rm tot}=0,2$ and $4$ channels. The whole $c_1$ versus $c_2$ parameter 
space can be divided into sub spaces with distinct symmetry properties of 
the densities of spinor components. We have the ferromagnetic phase for 
$c_1<0$ and $c_2>20c_1$,  anti-ferromagnetic phase for $c_2<0$ and  
$c_2<20c_1$, and cyclic phase for $c_1>0$ and $c_2>0$. In the ferromagnetic 
phase increasing magnetization lowers energy, whereas in the 
anti-ferromagnetic phase the lowest energy is attained for zero magnetization.   
Miscible configuration for five component densities of an SO-coupled spinor 
BEC is obtained for $c_2<20c_1$ and a phase separation is possible for 
$c_2>20c_1$. We find that for sufficiently strong spin-orbit coupling, the 
SO-coupled spinor condensate has atoms only in $m_f = 2$ and $m_f=-2$
states. Time-reversal-symmetry is preserved for states only in the 
anti-ferromagnetic phase, and can be broken in other phases. As parity is not 
a good quantum number, it is broken in all domains. We use the  numerical 
solution of the  generalized mean-field GP equation \cite{H_Wang} for this 
investigation.

The paper is organized as follows. In Sec.~\ref{Sec-II}, we describe the
coupled GP equations used to study the SO-coupled $f=2$ spinor BEC in a
quasi-1D trap. From an analytic  consideration of energy minimization, we 
predict the expected density  profiles for different set of parameters. 
Specifically, we predict the parameter space, where a phase separation 
can take place. We also predict different types of parity-breaking states. 
In Sec.~\ref{Sec-III}, we numerically study the SO-coupled spinor BEC in 
a quasi-1D trap. We identify the different types of symmetry-breaking states 
and phase-separated states in different parameter domains. We conclude by 
providing a summary of this study in Sec. \ref{Sec-IV}.

\section{Mean-field model for a SO-coupled BEC}
\label{Sec-II}  
Spin-orbit coupling can be generated for the hyper-fine states of neutral atoms 
by suitably controlling the atom-light interaction. The idea was realized 
experimentally by Lin {\it et al.} \cite{Lin} for two hyper-spin components of 
the $^{87}$Rb hyper-fine state 5S$_{1/2}$ employing two counter-propagating 
Raman lasers of wavelength $\lambda_r$ oriented at an angle $\beta_r$. This lead 
to the SO coupling with strength $\gamma = \hbar k_r/m$, where 
$k_r = 2\pi\sin(\beta_r/2)/\lambda_r$ and $m$ is the mass of an atom. 
However, here we consider the SO coupling among
the five spin components of the $f=2$ state, e.g.,  
$| f=2, m_f=2\rangle$, $| f=2, m_f=1\rangle$, $| f=2, m_f=0\rangle$, 
$| f=2, m_f=-1\rangle$ and $| f=2, m_f=-2\rangle$,
where   $m_f$ is the $z$ projection of $f$. By generalizing the method discussed
in Ref. \cite{lan}, this SO coupling among the five hyper-fine spin components 
can be generated by engineering a suitable atom-light interaction Hamiltonian 
as in Ref. \cite{Lin}.

In order to realize a quasi-1D SO-coupled spin-2 BEC along $x$ axis, we consider 
a trapping potential with angular frequencies along $y$ and $z$ axes much larger 
than that along $x$ axis. The resultant strong transverse confinement ensures 
that the dynamics is frozen along $y$ and $z$ axes. Then, the single-particle 
quasi-1D Hamiltonian of the system under the action of a strong transverse trap 
of angular frequencies $\omega_y$ and $\omega_z$ along $y$ and $z$ axes, 
respectively, can be written as \cite{Lin,Y_Zhang}
\begin{equation}\label{1x}
 H_0 = \frac{p_x^2}{2m}+\gamma p_x \Sigma_z + \Omega \Sigma_x,
\end{equation} 
where $p_x=-i\hbar \partial_x$ is the momentum operator along $x$ axis,
$\Omega$ is the Rabi frequency \cite{Lin,Linx}, and $\Sigma_z$ and $\Sigma_x$ 
are the irreducible representations of the $z$ and $x$ components of the 
spin-$2$ matrix, and are given by 
\begin{eqnarray}
\Sigma_z= \left( \begin{array}{ccccc}
2 & 0 & 0 & 0 & 0\\
0 & 1 & 0 & 0 & 0\\
0 & 0 & 0 & 0 & 0\\
0 & 0 & 0 & -1 & 0\\
0 & 0 & 0 & 0 & -2\end{array} \right), 
\end{eqnarray} 
\begin{eqnarray}
\Sigma_x= \left( \begin{array}
 {ccccc}
0 & 1 & 0 & 0 & 0\\
1 & 0 & \sqrt{\frac{3}{2}} & 0 & 0\\
0 & \sqrt{\frac{3}{2}} & 0 & \sqrt{\frac{3}{2}} & 0\\
0 & 0 & \sqrt{\frac{3}{2}} & 0 & 1\\
0 & 0 & 0 & 1 & 0\end{array} \right) .
\end{eqnarray}

\subsection{Mean-field model in a quasi-1D trap}
\label{Sec-IIA}
If the interactions among the atoms in the BEC are taken into account, in the 
Hartree approximation, using the single particle model Hamiltonian (\ref{1x}), a 
quasi-1D \cite{luca} spin-2 BEC can be described by the following set of five 
coupled mean-field  partial differential GP equations for the wave-function 
components  $\psi_j$  \cite{ueda,H_Wang}
\begin{align}
 i\hbar&\frac{\partial \psi_2}{\partial t} =
 \left( -\frac{\hbar^2}{2m}\frac{\partial^2}{\partial x^2}
 +V(x)+c_0\rho\right)\psi_2-2i\hbar\gamma
  \frac{\partial\psi_2}{\partial x}\nonumber\\ &+c_1\big(F_{-}
   \psi_1+2F_z\psi_2\big) 
  +\big({c_2}/{\sqrt{5}}\big)\Theta\psi_{-2}^*+ \Omega\psi_1, \label{gp-1}\\
i\hbar&\frac{\partial \psi_1}{\partial t} =
 \left( -\frac{\hbar^2}{2m}\frac{\partial^2}{\partial x^2}
 +V(x)+c_0\rho\right)\psi_1-i\hbar\gamma
  \frac{\partial\psi_1}{\partial x}  \nonumber\\
  & +c_1\big(\sqrt{3/2}F_{-}\psi_0+F_{+}\psi_2+F_z\psi_1\big) 
  -\big({c_2}/\sqrt{5}\big)\Theta\psi_{-1}^*\label{gp-2}\nonumber\\&
  + \Omega\big[\psi_2+\big(\sqrt{3}/2\big)\psi_0\big],
  \\
i\hbar &\frac{\partial \psi_0}{\partial t} = 
 \left( -\frac{\hbar^2}{2m}\frac{\partial^2}{\partial x^2}
 +V(x)+c_0\rho\right)\psi_0 +\frac{c_2}{\sqrt{5}}\Theta\psi_{0}^*  \nonumber\\
  &+\frac{\sqrt{6}}{2}c_1\big(F_{-} \psi_{-1}+F_{+}\psi_1\big) 
  + \frac{\sqrt{3}}{2}\Omega \big(\psi_1+\psi_{-1}\big),\label{gp-3}
  \\
 i\hbar &\frac{\partial \psi_{-1}}{\partial t} = 
 \left( -\frac{\hbar^2}{2m}\frac{\partial^2}{\partial x^2}
 +V(x)+c_0\rho\right)\psi_{-1}+i\hbar\gamma
  \frac{\partial\psi_{-1}}{\partial x} \nonumber\\
 &+c_1\big(\sqrt{3/2}F_{+}\psi_0+F_{-}\psi_{-2}-F_z\psi_{-1}\big) 
  -\big(c_2/\sqrt{5}\big)\Theta\psi_{1}^*\label{gp-4}\nonumber \\&
   + \Omega\big[\big({\sqrt{3}}/{2}\big)\psi_0+\psi_{-2}\big],
\\
 i\hbar &\frac{\partial \psi_{-2}}{\partial t} = 
 \left( -\frac{\hbar^2}{2m}\frac{\partial^2}{\partial x^2}
 +V(x)+c_0\rho\right)\psi_{-2}+2i\hbar\gamma
  \frac{\partial\psi_{-2}}{\partial x}\nonumber\\
  & +c_1\big(F_{+} \psi_{-1}-2F_z\psi_{-2}\big) 
  +\big(c_2/\sqrt{5}\big)\Theta\psi_{2}^*+ \Omega\psi_{-1},
\label{gp-5}
\end{align}
where $V(x) = m\omega_x^2x^2/2$ is the 1D harmonic trap, 
$c_0 = 2\hbar^2 (4a_2+3a_4)/(7m l_{yz}^2)$, 
$c_1 = 2\hbar^2 (a_4-a_2)/(7m l_{yz}^2)$, $c_2 = 2\hbar^2 (7a_0-10a_2+3a_4)/(7m l_{yz}^2)$,  
$a_0$, $a_2$ and $a_4$ are the $s$-wave scattering lengths in the total 
spin $f_{\mathrm{tot}} = 0,2$ and $4$ channels, respectively,    
$\rho_j = |\psi_j|^2$ with $j=2,1,0,-1,-2$ are the component densities, 
$\rho(x)= \sum_{j=-2}^2\rho_j$ is the total density, 
and $l_{yz} = \sqrt{\hbar/(m\omega_{yz})}$ with  $ \omega_{yz}=\sqrt{\omega_y\omega_z}$
is the oscillator length in the transverse $y-z$ plane and
\begin{align}
F_{+} = F_{-}^*&=2(\psi_2^*\psi_1+\psi_{-1}^*\psi_{-2})
+\sqrt{6}(\psi_1^*\psi_0
+\psi_0^*\psi_{-1}),\nonumber \\
F_z &= 2(|\psi_2|^2-|\psi_{-2}|^2) + |\psi_1|^2-|\psi_{-1}|^2,\nonumber \\
 \Theta &= \frac{2\psi_2\psi_{-2}-2\psi_1\psi_{-1}+\psi_0^2}{\sqrt{5}}.\nonumber 
\end{align} 
Here $\mathbf F = (F_x,F_y,F_z)$ is the spin density vector; $F_\pm =F_x\pm F_y.$  
The normalization condition is $\int_{-\infty}^\infty dx \rho(x)=N,$
where $N$ is the total number of atoms.
In order to transform  Eqs. (\ref{gp-1}) - (\ref{gp-5}) into dimensionless 
form, we use the scaled variables defined as
\begin{equation}
 \tilde{t} = \omega_x t,~\tilde{x} = \frac{x}{l_0},
 ~\phi_j(\tilde{x},\tilde{t}) = 
 \frac{\sqrt{l_0}}{\sqrt{N}}\psi_j(\tilde{x},\tilde{t}), 
\end{equation}
where $l_0=\sqrt{\hbar/(m\omega_{x}}$) is the oscillator length along $x$ axis.
Using these dimensionless variables, 
the coupled mean-field Eqs.~(\ref{gp-1}) - (\ref{gp-5}) in dimensionless form 
are
\begin{align} 
 i& \frac{\partial \phi_2}{\partial \tilde{t}} =
 \left( -\frac{1}{2}\frac{\partial^2}{\partial \tilde{x}^2}
 +\tilde{V}+\tilde{c}_0\tilde{\rho}\right)\phi_2-2i\tilde{\gamma}
  \frac{\partial\phi_2}{\partial \tilde{x}}
\nonumber\\
  & +\tilde{c}_1\big(\tilde{F}_{-} \phi_1+2\tilde{F}_{\tilde{z}}\phi_2\big) 
  +\big({\tilde{c}_2}/{\sqrt{5}}\big)\tilde{\Theta}\phi_{-2}^*\label{gp_s1}
 + \tilde{\Omega}\phi_1,\\
 i & \frac{\partial \phi_1}{\partial \tilde{t}} = 
 \left( -\frac{1}{2}\frac{\partial^2}{\partial \tilde{x}^2} 
  +\tilde{V}+\tilde{c}_0\tilde{\rho}\right)\phi_1  -i\tilde{\gamma}
  \frac{\partial\phi_1}{\partial \tilde{x}}\nonumber\\
 &+\tilde{c}_1\big(\sqrt{3/2} \tilde{F}_{-}\phi_0+\tilde{F}_{+}\phi_2+\tilde{F}_{\tilde{z}}\phi_1\big) 
  -\big(\tilde{c}_2/\sqrt{5}\big)\tilde{\Theta}\phi_{-1}^*\label{gp_s2}\nonumber \\&
 + \tilde{\Omega}\big[\phi_2+\big({\sqrt{3}}/{2}\big)\phi_0\big],\\
i&\frac{\partial \phi_0}{\partial \tilde{t}} =
 \left( -\frac{1}{2}\frac{\partial^2}{\partial \tilde{x}^2}
 +\tilde{V}+\tilde{c}_0\tilde{\rho}\right)\phi_0   +\frac{\tilde{c}_2}{\sqrt{5}}\tilde{\Theta}\phi_{0}^*\nonumber\\
  &+\frac{\sqrt{6}}{2}\tilde{c}_1\big(\tilde{F}_{-}
 \phi_{-1}+\tilde{F}_{+}\phi_1\big) 
  + \frac{\sqrt{3}}{2}\tilde{\Omega}(\phi_1+\phi_{-1}),\label{gp_s3}
  \\
 i&\frac{\partial \phi_{-1}}{\partial \tilde{t}} =
 \left( -\frac{1}{2}\frac{\partial^2}{\partial \tilde{x}^2}
 +\tilde{V}+\tilde{c}_0\tilde{\rho}\right)\phi_{-1}   +i\tilde{\gamma}
  \frac{\partial\phi_{-1}}{\partial \tilde{x}}
\nonumber\\
  & +\tilde{c}_1\big({\sqrt{3/2}}\tilde{F}_{+}\phi_0+\tilde{F}_{-}\phi_{-2}-\tilde{F}_{\tilde{z}}\phi_{-1}\big) 
  -\big({\tilde{c}_2}/{\sqrt{5}}\big)\tilde{\Theta}\phi_{1}^*\label{gp_s4}\nonumber\\&
 + \tilde{\Omega}\big[\big({\sqrt{3}}/{2}\big)\phi_0+\phi_{-2}\big],
  \\
 i&\frac{\partial \phi_{-2}}{\partial \tilde{t}} =
 \left( -\frac{1}{2}\frac{\partial^2}{\partial \tilde{x}^2}
 +\tilde{V}+\tilde{c}_0\tilde{\rho}\right)\phi_{-2}+2i\tilde{\gamma}
  \frac{\partial\phi_{-2}}{\partial \tilde{x}} \nonumber\\
  &+\tilde{c}_1\big(\tilde{F}_{+} \phi_{-1}-2\tilde{F}_{\tilde{z}}\phi_{-2}\big) 
  +\big({\tilde{c}_2}/{\sqrt{5}}\big)\tilde{\Theta}\phi_{2}^*
  + \tilde{\Omega}\phi_{-1}, 
\label{gp_s5}
\end{align}
where $\tilde{V} = \tilde{x}^2/2$, 
$\tilde{\gamma} = \hbar k_r/(m\omega_x l_0)$, 
$\tilde{\Omega} = \Omega/(\hbar \omega_x)$,
$\tilde{c}_0 = 2N (4a_2+3a_4)/(7l^2 _{yz})$,
$\tilde{c}_1 = 2N (a_4-a_2)/(7l^2 _{yz})$, 
$\tilde{c}_2 = 2N (7a_0-10a_2+3a_4)/(7l^2_{yz})$,
$\tilde{\rho}_j = |\phi_j|^2$ with $j=2,1,0,-1,-2$, 
and $\tilde{\rho} = \sum_{j=-2}^2|\phi_j|^2$
and
\begin{align}
\tilde{F}_{+} = \tilde{F}_{-}^*=& 2(\phi_2^*\phi_1+\phi_{-1}^*\phi_{-2})
+\sqrt{6}(\phi_1^*\phi_0 +\phi_0^*\phi_{-1}),\nonumber \\
\tilde{F}_{\tilde{z}} =& 2(|\phi_2|^2-|\phi_{-2}|^2) + |\phi_1|^2-|\phi_{-1}|^2,\nonumber \\
\tilde{\Theta} =& \frac{2\phi_2\phi_{-2}-2\phi_1\phi_{-1}+\phi_0^2}{\sqrt{5}}.\nonumber 
\end{align}
The normalization condition satisfied by $\phi_j$'s is
$
 \int_{-\infty}^{\infty} \tilde{\rho}(\tilde{x})d\tilde{x} = 1.
$
One of the aims in the present work is to find the ground state of an $f=2$ 
spinor condensate with a fixed magnetization, which is defined by 
\begin{eqnarray}
 {\cal M} = \int_{-\infty}^{\infty}\tilde{F}_{\tilde{z}}d\tilde x.
\label{magnetization}
\end{eqnarray}
Depending on the values of $\tilde c_1$ and $\tilde c_2$ the system in the absence of
magnetic field and SO coupling can have a variety of ground states
\cite{ueda}. For the sake of simplicity of notations, we will represent the 
dimensionless variables without tilde in the rest of the paper.

\subsection{Uniform BEC: Analytic Consideration}
\label{Sec-IIB}

The energy of a uniform (trapless) spinor BEC in the presence of SO coupling and 
magnetic field is 
\cite{ueda,H_Wang}
\begin{align}
 E &=  N\int_{-\infty}^{\infty} \Big\{\frac{1}{2}\sum_{j=-2}^2\left|\frac{d\phi_j}{dx}\right|^2
   -i\gamma\sum_{j=-2}^2j\phi_j^*\frac{d\phi_j}{dx}
  \nonumber\\
  &+ \big( {c_0}\rho^2 + {c_1}|\mathbf F|^2+{c_2}|\Theta|^2\big)/2
 +\Omega\big[\phi_2^*\phi_1+\phi_1^*\phi_2\nonumber\\
   &+\phi_{-1}^*\phi_{-2}+\phi_{-2}^*\phi_{-1}+\big({\sqrt{3}}/{2}\big)\big(\phi_1^*\phi_0+\phi_0^*\phi_1+\nonumber \\
&\phi_0^*\phi_{-1}+\phi_{-1}^*\phi_0\big) \big]\Big\}dx.
  \label{energy}
\end{align}

First we consider the formation of spatially-separated non-overlapping 
(phase-separated) states from a consideration of energy minimization. As in 
the case of a $f=1$ spinor BEC \cite{sandeep}, the energy term proportional 
to $c_0$ in Eq. (\ref{energy}) can not lead to a phase separation as it 
contains terms $Nc_0 \int(\rho_j^2/2 + \rho_j'^2/2+\rho_j\rho_j')dx$, 
where $j,j' = 2, 1, 0, −1, -2$ and $j\ne j'$, and hence corresponds 
to a situation where inter- and intra-species interactions are of equal 
strength. The situation 
is analogous to a binary BEC with a $a_{12} = \sqrt{a_{11} a_{22}}$ ,
where $a_{11}$ and $a_{22}$ are intra-species and $a_{12}$ the 
inter-species scattering lengths. Such a binary BEC has equal
strengths of inter- and intra-species nonlinearities and is
always miscible in the presence of a 1D harmonic trap \cite{Ao,Gautam}. The interaction 
energy of the $f=2$ spinor condensate in the absence of SO coupling and magnetic field ($\gamma=\Omega=0$)
can be written
as
\begin{align}\label{aaa}
&E_{\rm int} = N\int\Big\{\frac{c_1}{2}\Big[4\rho_2^2+\rho_1^2+\rho_{-1}^2+4\rho_{-2}^2\nonumber\\
&+           6\rho_0(\rho_1+\rho_{-1})+8\sqrt{\rho_1\rho_2\rho_{-1}\rho_{-2}}\cos(\theta_{-1-2}-\theta_{21})\nonumber\\
 &+12\rho_0\sqrt{\rho_1\rho_{-1}}\cos(\theta_{0-1}-\theta_{10})+4\sqrt{6}\Big(\rho_{-1}\sqrt{\rho_{-2}\rho_0}\nonumber\\
       &\times  \cos(\theta_{-1-2}-\theta_{0-1})+\sqrt{\rho_{-1}\rho_{-2}\rho_0\rho_1}\cos(\theta_{-1-2}-\theta_{10})\nonumber\\
       &+
          \sqrt{\rho_2\rho_1\rho_0\rho_{-1}}\cos(\theta_{21}-\theta_{0-1})+\rho_1\sqrt{\rho_2\rho_0}
 \cos(\theta_{21}-\theta_{10})\Big)\nonumber\\
       &+8\rho_2\rho_1+8\rho_{-1}\rho_{-2}-4\rho_2\rho_{-1}-8\rho_2\rho_{-2}-2\rho_1\rho_{-1}\nonumber\\
&-4\rho_1\rho_{-2}\Big]+\frac{c_2}{10}\Big[ 4\rho_2\rho_{-2}+4\rho_1\rho_{-1}-8\sqrt{\rho_2\rho_1\rho_{-2}\rho_{-1}}\nonumber\\
&\times \cos(\theta_{21}-\theta_{-1-2})+
        4\rho_0\left(\sqrt{\rho_2\rho_{-2}}\cos(\theta_{20}-\theta_{0-2})\right.\nonumber\\
 &\left.-\sqrt{\rho_1\rho_{-1}}\cos(\theta_{10}-\theta_{0-1})\right)+\rho_0^2\Big]\Big\}dx,
\end{align}
where the wave-function component $\phi_j$ is written as $\phi_j=\sqrt{\rho_j}\exp(i\theta_j)$ with $\theta_j$ the phase. The  phase difference between the $i$th and $j$th 
components is written as  $\theta_{ij} = \theta_i -\theta_j$. 

To understand the phase separation and spontaneous symmetry breaking of the states, 
we consider the stationary eigenvalue problem of the lowest-energy state of 
 a uniform non-interacting system with SO coupling while Eqs. (\ref{gp_s1}) - (\ref{gp_s5}) become
\begin{equation}\label{dd}
E \phi_j(x)= N\left[-\frac{1}{2}\frac{\partial^2}{\partial x^2} -ij\gamma
  \frac{\partial}{\partial x}   \right]\phi_j(x).
\end{equation}
The two independent solutions of Eq. (\ref{dd})
for the lowest-energy state are $\phi_{\pm 2}=\alpha_{\pm 2} \exp(\mp 2 i \gamma x) $ with normalization $|\alpha_2|^2+|\alpha_{-2}|^2=1$ and $
\phi_{\pm 1}= \phi_{\pm 0}=0$ with energy   $E\equiv E_{\mathrm{min}}=-2N\gamma^2$. 
The components $j=\pm 1,0$ have higher energies. In the presence of a trap and interactions, 
in the actual numerical calculation, for a sufficiently large SO coupling $\gamma$ only the 
components $j=\pm 2$ survive. Hence, the analytic solutions of Eq. (\ref{dd}) is very useful 
to understand many features of the actual numerical solution. It is clear from Eq. (\ref{dd}) 
that these plane-wave solutions will lead to smooth density profiles in the presence  
of a trap while their real and imaginary parts will, 
in general, have oscillating behavior.
In the case, when the two components $j=\pm 2$ survive,  the interaction energy (\ref{aaa})
becomes 
\begin{align}
E_{\rm int} = 2N\int \Big[c_1 (\rho_2+\rho_{-2})^2 
      +\frac{c_2-20c_1}{5} \rho_2\rho_{-2}\Big]
dx. \label{bbx}
\end{align}
In Eq. (\ref{bbx}), only the product term $\rho_2\rho_{-2}$ controls the phase separation 
between components $\pm 2$. A repulsive (positive) product term will facilitate a
phase separation, as the energy 
can then be minimized by reducing the overlap between the  components. 
This will happen for  $ c_2>20c_1.$
 
In the presence of  SO coupling the Hamiltonian is invariant under time reversal  
${\cal T}$ but not under parity. Hence, parity is not a good quantum number. 
Different types of simple parity-breaking states are found. The non-degenerate 
states should possess time-reversal symmetry. However, a pair of degenerate
states, which transform into each other when operated upon by ${\cal T},$
break time-reversal symmetry. For a sufficiently large $\gamma$, when only
the components $j=\pm2$ survive, the time reversal symmetry is broken
for the phase-separated profiles, whereas the miscible profiles preserve
time-reversal symmetry.  Next we consider different types of symmetry-breaking states.

First we consider different overlapping  parity-breaking, nevertheless time-reversal-symmetric,  states.
The parity property of $\phi_{\pm 2}$ and the relations between the real and imaginary parts, denoted
by ${\cal R}$ and ${\cal I}$ respectively, of $\phi_2$ and $\phi_{-2}$ for some of these
parity-breaking and time-reversal-symmetric states are listed in Table \ref{table-1}.
\begin{table*}[ht]
\begin{center}
\centering
    \begin{tabular}{ | >{\centering\arraybackslash}m{0.8cm} | >{\centering\arraybackslash}m{0.8cm} 
| >{\centering\arraybackslash}m{8cm} | >{\centering\arraybackslash}m{7cm} | @{}m{0cm}@{}}
    \hline
    $\alpha_2$ &  $\alpha_{-2}$ &  Parity property of $\phi_{\pm 2}$ & 
    Relation between $\phi_2$ and $\phi_{-2}$\\ \hline
    $\frac{\pm 1}{\sqrt{2}}$ & $\frac{\pm 1}{\sqrt{2}}$ & ${\cal R} [\phi_{\pm 2}(x)]=
  {\cal R} [\phi_{\pm 2}(-x)]$,  
 ${\cal I} [\phi_{\pm 2}(x)]=-
  {\cal I} [\phi_{\pm 2}(-x)]$  &  $
 {\cal R} [\phi_{2}(x)]=
  {\cal R} [\phi_{- 2}(x)], $ $
 {\cal I} [\phi_{ 2}(x)]=-
  {\cal I} [\phi_{- 2}(x)]$ &\\[10pt] \hline 
    $\frac{\pm 1}{\sqrt{2}}$ & $\frac{\mp 1}{\sqrt{2}}$ & ${\cal R} [\phi_{\pm 2}(x)]=
  {\cal R} [\phi_{\pm 2}(-x)], $ $
 {\cal I} [\phi_{\pm 2}(x)]=-
  {\cal I} [\phi_{\pm 2}(-x)]$  &$
 {\cal R} [\phi_{2}(x)]=-
  {\cal R} [\phi_{- 2}(x)], $ $
 {\cal I} [\phi_{ 2}(x)]=
  {\cal I} [\phi_{- 2}(x)]$ &\\[10pt] \hline
    $\frac{\pm i}{\sqrt{2}}$ & $\frac{\pm i}{\sqrt{2}}$ & ${\cal R} [\phi_{\pm 2}(x)]=
-  {\cal R} [\phi_{\pm 2}(-x)], $ $
 {\cal I} [\phi_{\pm 2}(x)]=
  {\cal I} [\phi_{\pm 2}(-x)]$ &  $
 {\cal R} [\phi_{2}(x)]=-
  {\cal R} [\phi_{- 2}(x)], $ $
 {\cal I} [\phi_{ 2}(x)]=
  {\cal I} [\phi_{- 2}(x)]$ &\\[10pt]
    \hline
$\frac{\pm i}{\sqrt{2}}$ & $\frac{\mp i}{\sqrt{2}}$ & ${\cal R} [\phi_{\pm 2}(x)]=
-  {\cal R} [\phi_{\pm 2}(-x)], $ $
 {\cal I} [\phi_{\pm 2}(x)]=
  {\cal I} [\phi_{\pm 2}(-x)]$ & $
 {\cal R} [\phi_{2}(x)]=
  {\cal R} [\phi_{- 2}(x)], $ $
 {\cal I} [\phi_{ 2}(x)]=-
  {\cal I} [\phi_{- 2}(x)]$ &\\[10pt]
\hline
$\frac{ i}{\sqrt{2}}$ & $\frac{\pm 1}{\sqrt{2}}$ & ${\cal R} [\phi_{ 2}(x)]=
-  {\cal R} [\phi_{ 2}(-x)], $
${\cal R} [\phi_{ -2}(x)]=
  {\cal R} [\phi_{- 2}(-x)], $
${\cal I} [\phi_{ 2}(x)]=
  {\cal I} [\phi_{ 2}(-x)], $
${\cal I} [\phi_{ -2}(x)]=
  -{\cal I} [\phi_{- 2}(-x)] $
 & $
 {\cal R} [\phi_{2}(x)]=\pm
  {\cal I} [\phi_{-2}(x)],$ $
 {\cal I} [\phi_{2}(x)]=\pm 
  {\cal R} [\phi_{- 2}(x)]$ &\\[10pt]
\hline
$\frac{\pm 1}{\sqrt{2}}$ & $\frac{i}{\sqrt{2}}$ & ${\cal R} [\phi_{ 2}(x)]=
  {\cal R} [\phi_{ 2}(-x)], $
${\cal R} [\phi_{ -2}(x)]=-
  {\cal R} [\phi_{- 2}(-x)], $
${\cal I} [\phi_{ 2}(x)]=-
  {\cal I} [\phi_{ 2}(-x)], $
${\cal I} [\phi_{- 2}(x)]=
  {\cal I} [\phi_{- 2}(-x)] $& $
 {\cal R} [\phi_{2}(x)]=\pm
  {\cal I} [\phi_{-2}(x)],$ $
 {\cal I} [\phi_{2}(x)]=\pm 
  {\cal R} [\phi_{- 2}(x)]$ &\\[10pt]
\hline
    \end{tabular}
\caption{Parity-breaking states with different choices of $\alpha_{\pm 2}$. The
third column defines the parity of the real and imaginary parts of $\phi_{\pm 2}$, whereas
the fourth column shows the relations between the real and imaginary parts of $\phi_2$
and $\phi_{-2}$.}
\label{table-1}
\end{center}
\end{table*}
In these examples the real and imaginary parts of the wave function 
may have definite parity, but not the total wave function, as parity is not a good quantum number.
 However, no such simple relation is obtained for a general $\alpha_{\pm 2}$, 
where neither the real part nor the imaginary part of the wave-function components have a definite parity. Also, similar symmetry-breaking states are expected for the $j=\pm 1$ component states when they are nonzero. 
These types of symmetry-breaking states were found in the
actual numerical calculation in the presence of trap and interaction terms.

Now we consider some examples of time-reversal symmetry-breaking states in the presence of SO  coupling. 
These states are phase-separated (non-overlapping).
There could be a complete phase separation between the $j=\pm 2$ components when the two components symmetrically move to two sides of $x=0$. In that case, suppose the components  $j=\pm 2$ are centered at $x=\pm x_0$, then there will be no definite parity    of the real and imaginary parts, but one can have properties, such as,   $  {\cal R} [\phi_{ 2}(x-x_0)] = \pm  {\cal R} [\phi_{ -2}(x+x_0)]$, 
  $  {\cal I} [\phi_{ 2}(x-x_0)] = \mp  {\cal I} [\phi_{ -2}(x+x_0)]$,  or 
 $  {\cal R} [\phi_{ 2}(x-x_0)] = \pm  {\cal I} [\phi_{ -2}(x+x_0)]$,
 $  {\cal R} [\phi_{ -2}(x+x_0)] = \pm  {\cal I} [\phi_{ 2}(x-x_0)]$,  etc. In these cases 
the densities  break the symmetry of the trapping potential: $\rho_j(x)\ne \rho_j(-x).$
 There could be another type of phase separation where one of the components, say $j=-2$, stays at the middle and the other component
$j=2$  
breaks into two parts and stay symmetrically on both sides of origin. 
In this case, the real and imaginary parts of the middle component ($j=-2$) have 
opposite parities,  and the real and imaginary parts of the outer component ($j=2$) 
either map into each other or have  opposite parities: $
 {\cal R} [\phi_{2}(x)]=\pm
  {\cal I} [\phi_{2}(-x)],$ or   $
 {\cal R} [\phi_{2}(x)]=\pm 
  {\cal R} [\phi_{ 2}(-x)]$  and $
 {\cal I} [\phi_{2}(x)]=\mp 
  {\cal I} [\phi_{ 2}(-x)]$.
In these cases, the symmetry of the trapping potential is reflected in the densities:
 $\rho_j(x)= \rho_j(-x).$

For a moderate SO coupling, in the trapped system $\phi_0=0$  and interesting conclusions can be reached 
analytically in such a case. 
In the limit $|\phi_0|\rightarrow 0$,
the interaction energy is
\begin{align}
&E_{\rm int} = N\int\Big\{\frac{c_1}{2}\big[4(\rho_2+\rho_1+\rho_{-1}+\rho_{-2})^2 -3(\rho_1+\rho_{-1})^2\nonumber\\
&+8\sqrt{\rho_2\rho_{-2}\rho_{1}\rho_{-1}}\cos(\theta_{-1-2}-\theta_{21})-12\rho_2\rho_{-1}\nonumber\\
       &-16\rho_2\rho_{-2}-4\rho_1\rho_{-1}-12\rho_1\rho_{-2}\big]+\frac{c_2}{10}\big[ 4\rho_2\rho_{-2}\nonumber\\
&+4\rho_1\rho_{-1}-8\sqrt{\rho_2\rho_{-2}\rho_{1}\rho_{-1}}\cos(\theta_{21}-\theta_{-1-2})\big]\Big\}
dx. \label{bbb}
\end{align}
{ The {\it positive} terms in Eq. (\ref{bbb})
involving a product of different density components should 
enhance a phase separation, as they can be minimized by making the overlap between the component wave functions
zero. For the extremum values of $\cos (\theta_{-1-2}-\theta_{21})=\pm 1,$
the energy of  Eq. (\ref{bbb}) can be written as

\begin{align}
&E_{\rm int} = \frac{N}{2}\int\Big\{4c_1(\rho_2+\rho_1+\rho_{-1}+\rho_{-2})^2  -12c_1\rho_2\rho_{-1} \nonumber\\& -12c_1\rho_1\rho_{-2}
+\frac{4}{5}\big(c_2-{20c_1}\big) \big[ \sqrt{\rho_2\rho_{-2}}\mp \sqrt{\rho_1\rho_{-1}}\big]^2 
\nonumber\\
       &-3c_1(\rho_1-\rho_{-1})^2
\mp 24c_1\sqrt{\rho_2\rho_{-2}\rho_1\rho_{-1}
}
 \Big\}
dx. \label{ccc}
\end{align}
Let us in addition consider  zero magnetization: $\int dx \rho_j=\int dx\rho_{-j}, 
j\ne 0$. 
For  $c_1>0, c_2>20c_1$,  the crossed terms in density involving $\rho_j$ and $\rho_{-j}$ are positive 
and those involving $\rho_{\pm 2}$ and $\rho_{\mp 1}$ are negative. Hence a stable state with energy minimization will correspond to a phase separation between components $\pm 2$ and between $\pm 1$ 
while maintaining overlap between components $2$ and $-1$ and between $-2$ and $1$. In case of this phase separation 
the term $\mp 24c_1\sqrt{\rho_2\rho_{-2}\rho_1\rho_{-1}}$ will contribute zero.
Similarly, for   $c_1>0$ and $c_2<20c_1$, the dominating contribution of the  terms in Eq. (\ref{ccc}) involving a product of densities are negative, and 
can be minimized by increasing the overlap between components and one can never have a phase separation.
For $c_1<0$ and $c_2>20c_1$, the dominating contribution of the  terms in Eq. (\ref{ccc}) involving a product of densities are 
  positive,  
{and can be minimized by accommodating all the atoms in phase-separated $m_f=2$ and $-2$ components.
All the atoms in $m_f=2$ and $-2$ components also ensure the minimum contribution from the $-3Nc_1\int(\rho_1-\rho_{-1})^2/2dx$
term.}  
 Finally, for  $c_1<0, c_2<20c_1$, the crossed terms in density involving $\rho_j$ and $\rho_{-j}$ are negative and those involving $\rho_{\pm 2}$ and $\rho_{\mp 1}$ are positive. 
Hence if  a phase separation occurs it will be  between components $2$ and $-1$ and between $-2$ and $1$
while maintaining overlap between components $\pm 2$ and between $\pm 1$. 
However, a consideration of minimization of the repulsive contribution 
${-3Nc_1\int(\rho_1-\rho_{-1})^2/2dx}$ to energy (\ref{ccc}) for overlapping 
$\pm \rho_j$ requires $\rho_1=\rho_{-1}=0$, which  will exclude the possibility of 
a phase separation.

There are several known phases of this spin-2 system.
For $c_1<0, c_2>0$, the state of largest magnetization corresponds to the lowest-energy state
and such states are termed ferromagnetic. Even in the absence
of SO coupling these states violate time-reversal symmetry.
 For $c_1>0, c_2<0$, the state of zero magnetization has the lowest energy corresponding to the anti-ferromagnetic, or polar, or nematic phase. These states with ${\cal M}=0$
preserve time-reversal symmetry in the absence
as well as presence of SO coupling.
The ferromagnetic and anti-ferromagnetic phases also extend to the domain $-$  $c_1<0, c_2<0$ $-$ however, separated by the line 
$c_2=20c_1.$ For $c_1>0,c_2>0$ neither ferromagnetic nor anti-ferromagnetic property prevails and a new phase termed cyclic  emerges. A separation of phase is expected for ferromagnetic material and not for 
anti-ferromagnetic material. In this domain of cyclic phase, the time-reversal symmetry is
broken in the absence of SO coupling. In the presence of a sufficiently strong SO coupling, when only the components
$j = \pm 2$ survive, the time-reversal symmetry is broken
in the phase-separated domain, i.e.,  $c_2 > 20c_1$, while it
is preserved in the miscible domain, i.e., $ c_2 < 20c_1.$
 It is interesting that the present analytic discussion from a consideration of energy minimization could predict the line $c_2=20c_1$ separating the ferromagnetic and anti-ferromagnetic phases corresponding to phase-separated and overlapping states, respectively.

In the following section,
by numerically solving the coupled Eqs.~(\ref{gp_s1}) - (\ref{gp_s5}), we will
show that for {$c_2>20c_1$}, $|\phi_0|\rightarrow 0$ is sufficient but not necessary condition for a 
phase separation.

\section{Numerical solution of the coupled GP equation}
\label{Sec-III}
We study the ground state structure of the spinor BEC by solving 
the coupled Eqs.~(\ref{gp_s1}) - (\ref{gp_s5}) numerically using a split-time-step
Crank-Nicolson method
\cite{Muruganandam,H_Wang}. The spatial and time steps employed in the
present work are $\delta x = 0.05$ and $\delta t= 0.000125$. In order to 
find the ground state, we employ 
imaginary-time propagation. The imaginary time propagation neither conserves 
the normalization nor the magnetization as the (imaginary) time evolution operator is not unitary. 
To fix the  normalization,  and {consequently} preserve magnetization, we suggest 
the following approach \cite{Bao}.

\subsection{Calculation of the normalization constants} 

The minimization of energy given by Eq. (\ref{energy}) under the constraints
of fixed normalization (=1) and magnetization ($\cal M$) can be implemented by minimizing the functional
\begin{equation}
K = E -\mu \left(\int \sum_{j=-2}^2 |\phi_j|^2dx-1\right) -\lambda\left(\int F_z dx-{\cal M}\right),
\end{equation}
where $\mu$ and $\lambda$ are the Lagrangian multipliers and are functions
of $\phi_j$'s. The imaginary time
equivalent of  Eqs.~(\ref{gp_s1}) - (\ref{gp_s5}) using this functional can be written as 
\begin{equation}
-\frac{\partial\phi_j(x,\tau)}{\partial \tau} = \frac{\delta E}{\delta \phi_j^*(x,\tau)} - (\mu+j\lambda)\phi_j(x,\tau).
\label{cngf}
\end{equation}
This coupled set of equations is also termed as continuous normalized gradient
flow equations \cite{Cngf} for $f=2$ spinor BEC. Applying the first order time 
splitting to Eqs. (\ref{cngf}), we break up this equation into two parts
\begin{eqnarray}
-\frac{\partial\phi_j(x,\tau)}{\partial \tau} &=& \frac{\delta E}{\delta \phi_j^*(x,\tau)},\\
-\frac{\partial\phi_j(x,\tau)}{\partial \tau} &=& - (\mu+j\lambda)\phi_j(x,\tau),
\label{split_eq_2}
\end{eqnarray}
which have to be solved one after the other. The solution of Eq. (\ref{split_eq_2}) at $\tau=\tau+\delta \tau$
is analytically known:
\begin{align}
\phi_j(x,\tau+\delta \tau)\equiv & d_j\phi_j(x,\tau) \nonumber \\
= &\exp\biggr[{\int_{\tau}^{\tau+\delta\tau}(\mu+j\lambda)d\tau}\biggr]\phi_j(x,\tau),
\end{align}   
Using this definition of $d_j$, one can derive the following relations  \cite{Bao}:
\begin{eqnarray}
d_1d_{-1} &=& d_0^2,\label{norm_const1}\\
d_2d_{-2} &=& d_0^2,\\
d_2d_{-1}^2 &=& d_0^3.
\label{norm_const3}
\end{eqnarray}
Now, the constraints on norm and magnetization can be written in terms of the normalization
constants $N_j$ of the wave-function components as
\begin{flalign}
d_2^2N_2+d_1^2N_1+d_0^2N_0+d_{-1}^2N_{-1}+d_{-2}^2N_{-2} = 1\label{lv-1},\\
2d_2^2N_2+d_1^2N_1-d_{-1}^2N_{-1}-2d_{-2}^2N_{-2} = {\cal M},
\label{norm_mag}
\end{flalign}
where $N_j=\int |\phi_j(x,\tau)|^2dx$.
Equations (\ref{norm_const1}) - (\ref{norm_mag}) lead to the following set of
non-linear algebraic equations
\begin{flalign}
u^4N_2+vu^3N_1+v^2u^2N_0+v^3uN_{-1}+v^4N_{-2} = 1,\label{nl1}\\
2u^4N_2+vu^3N_1-v^3uN_{-1}-2v^4N_{-2} = {\cal M},\label{nl2}
\end{flalign}
where $u = d_1^2$ and $v = d_0^2$.
We use Newton-Raphson method \cite{num_recipies} for non-linear system of equations to solve
Eqs.~(\ref{nl1}) - (\ref{nl2}) after each iteration in imaginary time 
to determine $d_1$ and $d_0$ and hence the remaining
projection operators using Eqs. (\ref{norm_const1}) - (\ref{norm_const3}).

\begin{figure}[!t]
\begin{center}
\includegraphics[clip,width=.8\linewidth,clip]{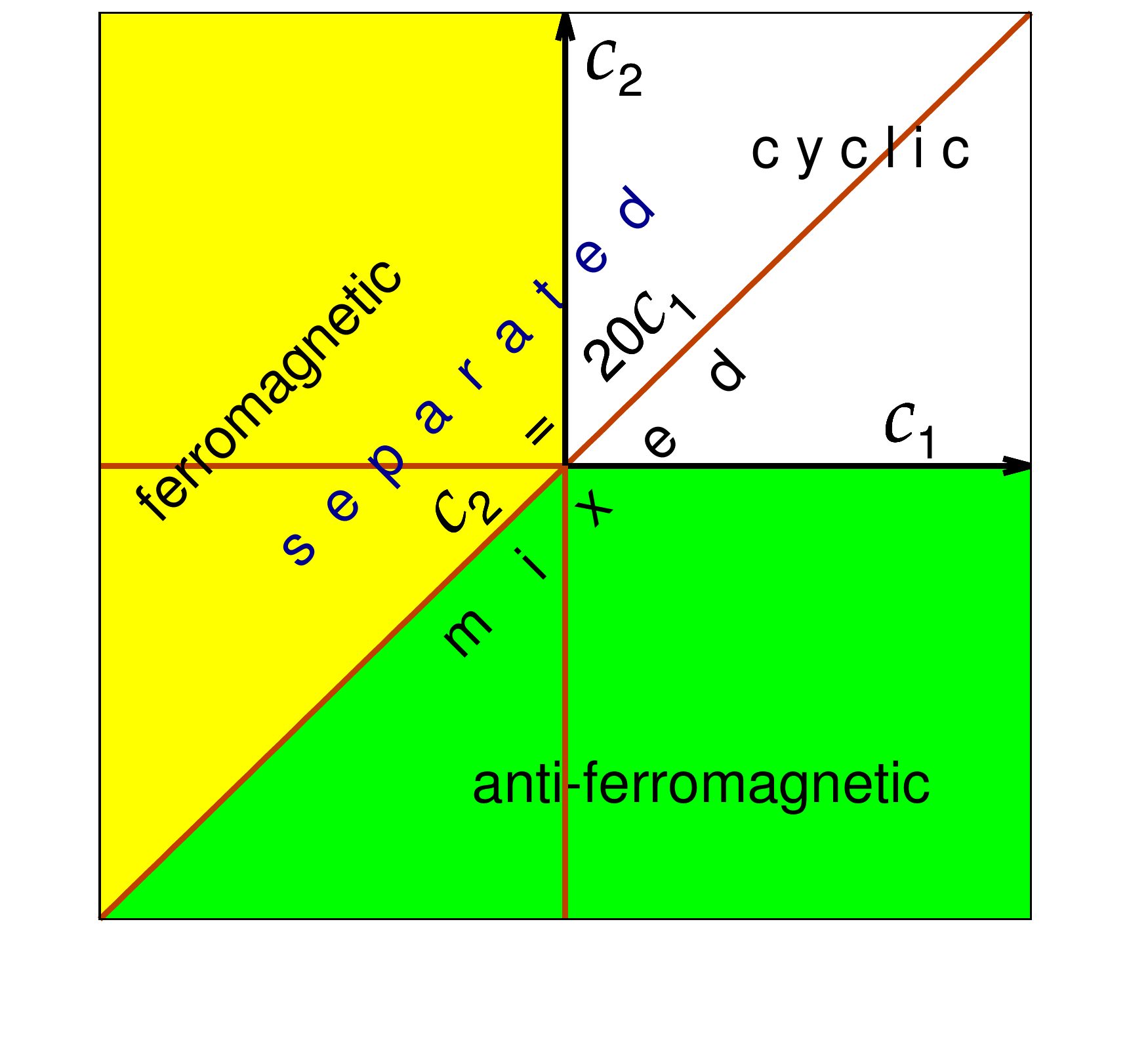} 

\caption{(Color online) The $c_2-c_1$  phase plot illustrating ferromagnetic, anti-ferromagnetic, and cyclic phases. The $c_2=20c_1$ line separating the ferromagnetic and anti-ferromagnetic phases as obtained from the present analytic consideration is shown. Separated phase is possible above this line and  miscible phase  below this line.}
\label{fig-0} \end{center}
\end{figure}

\begin{figure}[!t]
\begin{center}
\includegraphics[trim = 5mm 0mm 4cm 0mm, clip,width=.49\linewidth,clip]{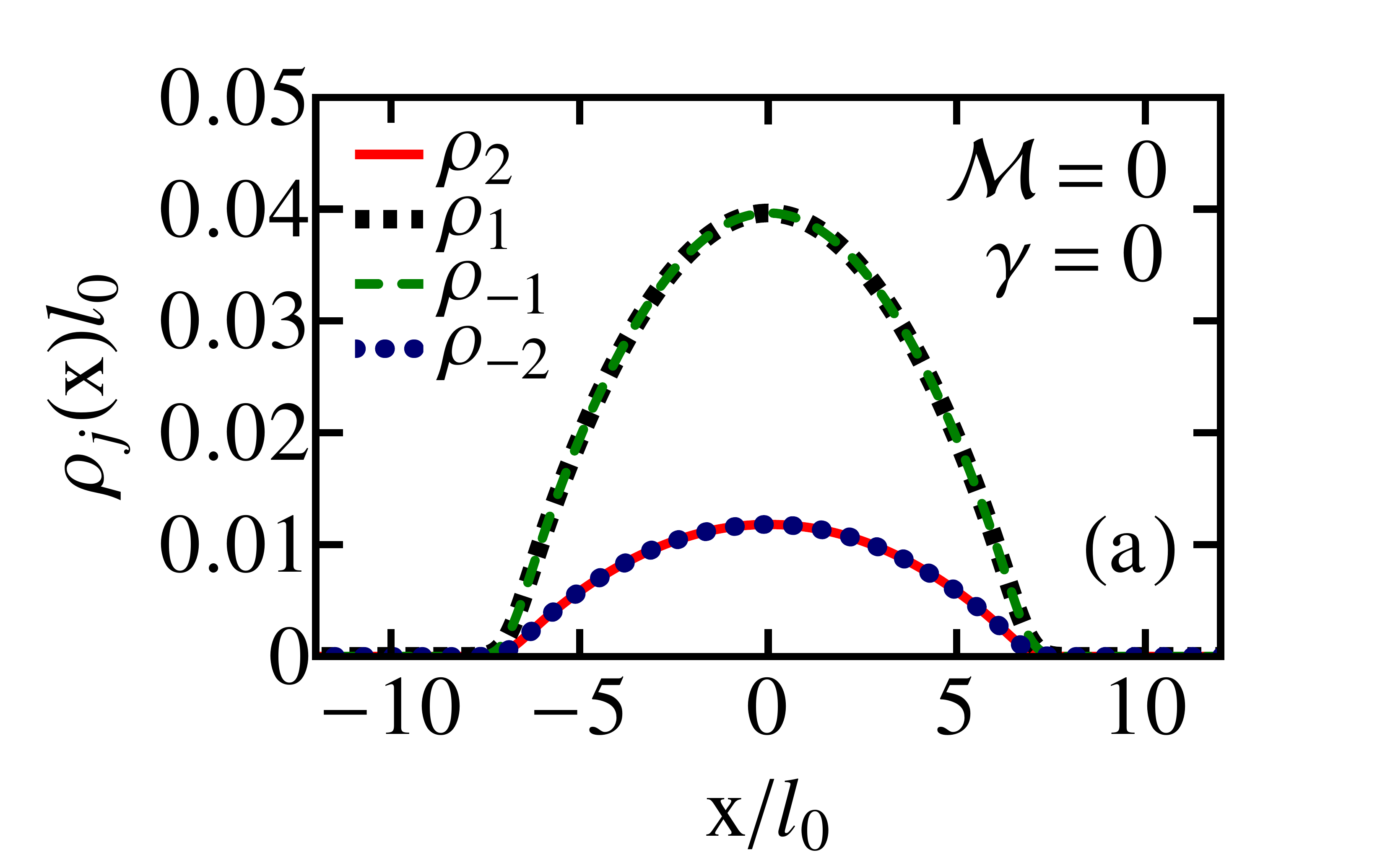}
\includegraphics[trim = 5mm 0mm 4cm 0mm, clip,width=.49\linewidth,clip]{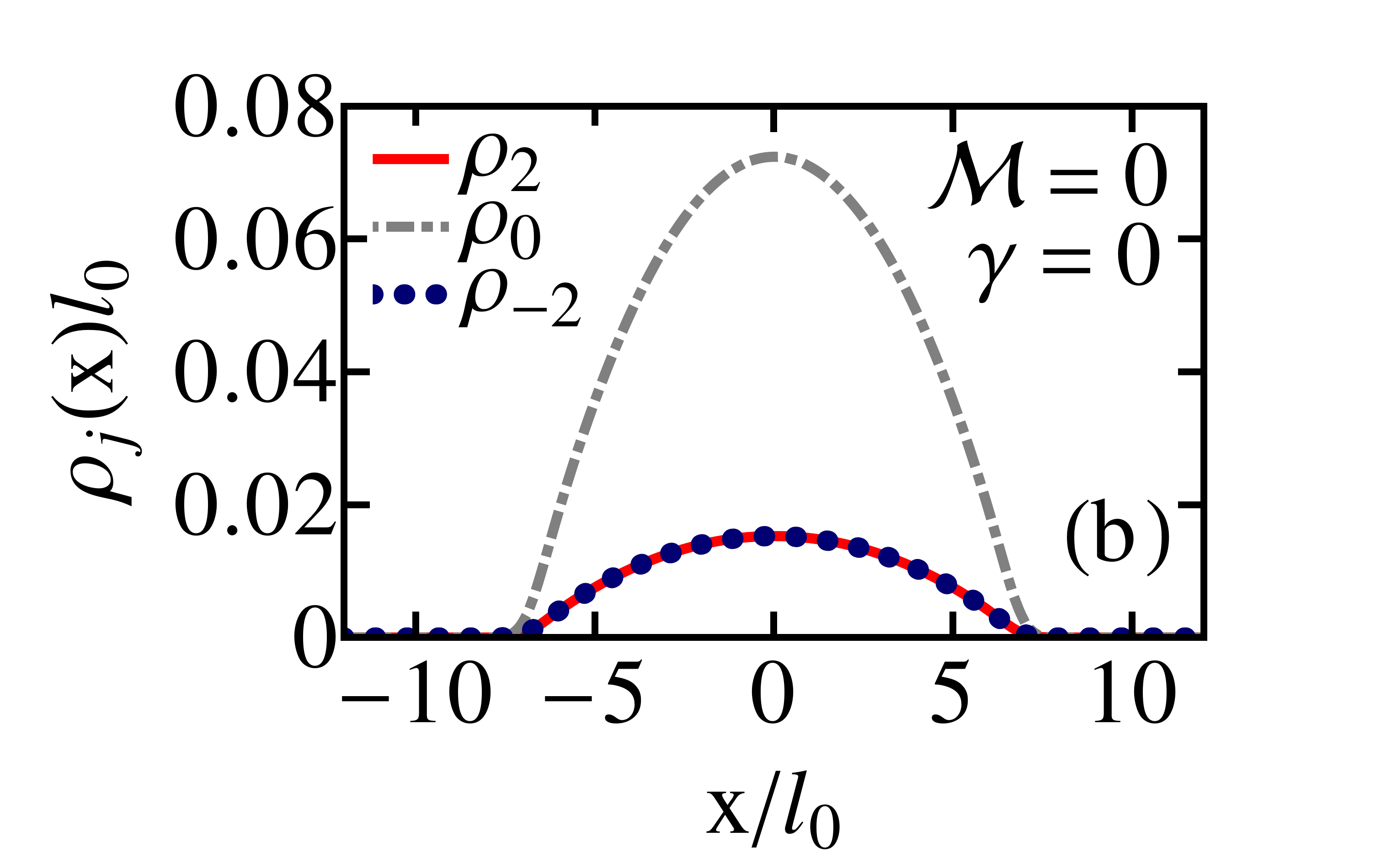}
\includegraphics[trim = 5mm 0mm 4cm 0mm, clip,width=.49\linewidth,clip]{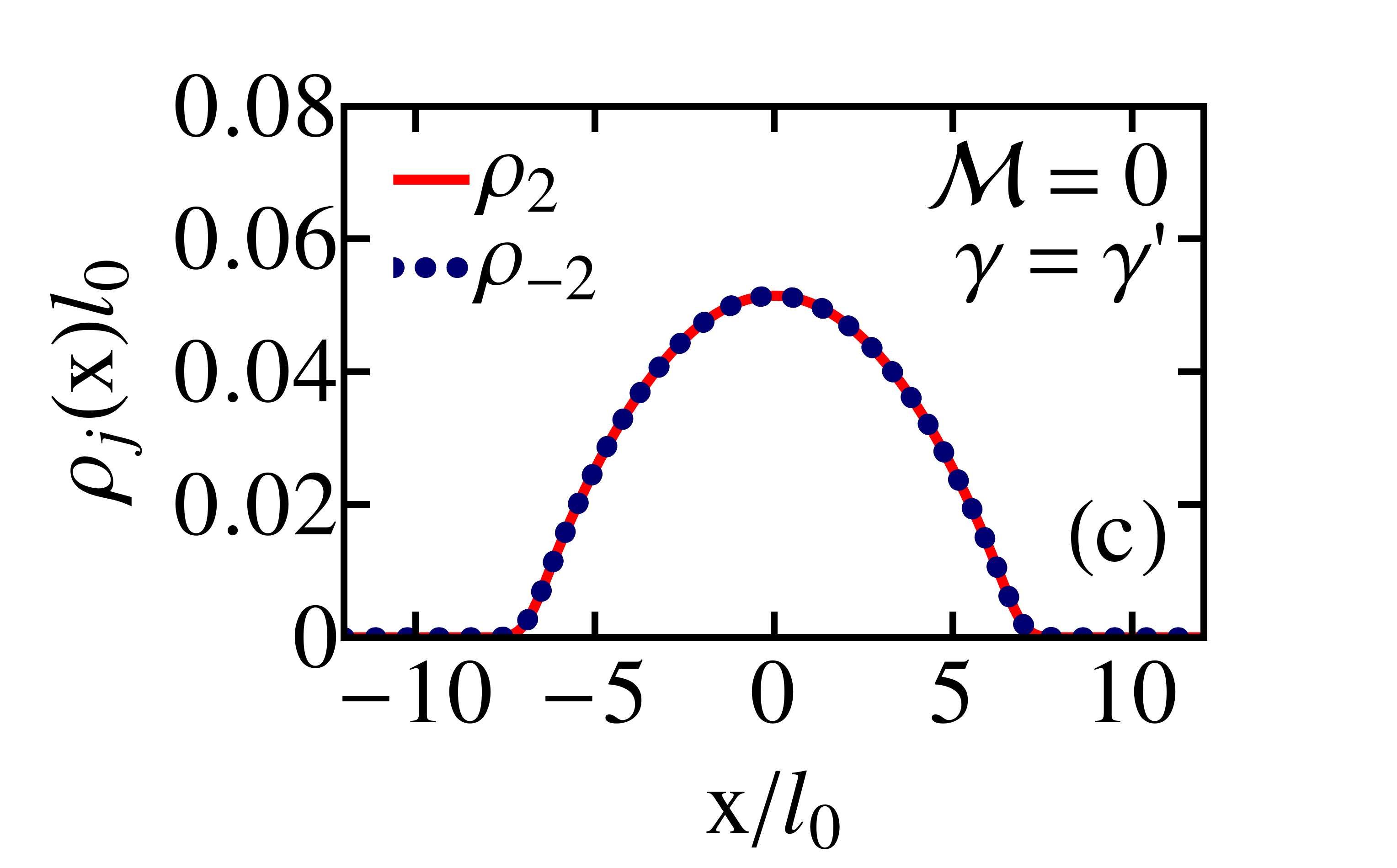}
\includegraphics[trim = 5mm 0mm 4cm 0mm, clip,width=.49\linewidth,clip]{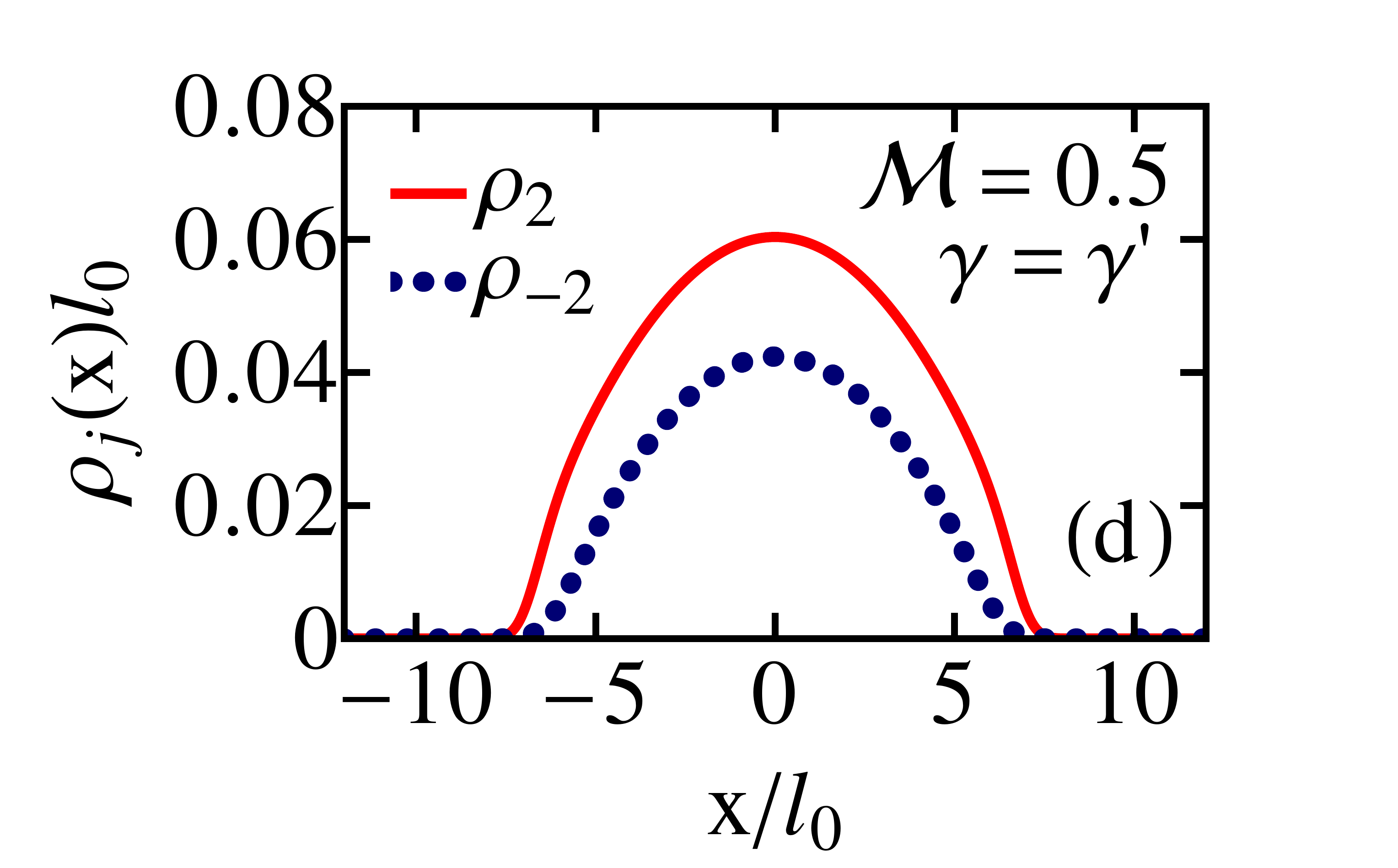}
\caption{(Color online) Component densities  of a $^{23}$Na spinor BEC of
$10000$ atoms for $\Omega=0$ and $c_0 = 242.97, c_1 = 12.06 > 0,$ and $c_2 =-
13.03 < 20c_1$. The parameters used are 
(a) $\gamma={\cal M}=0$, (b)  $\gamma={\cal M}=0$, (c)  $\gamma=\gamma', {\cal M}=0$, 
and (d)  $\gamma=\gamma', {\cal M}=0.5$; here $\gamma'$ can have any arbitrary real value including 0,
and hence denotes any arbitrary strength of SO coupling. 
Time-reversal symmetry is broken in (d). In this and following similar figures all quantities are dimensionless and only the nonzero density components are shown.}
\label{fig-1} \end{center}
\end{figure}

\subsection{Numerical Results}

We consider $10000$ atoms of $^{23}$Na (in $f=2$ hyper-fine spin state) in a trapping potential 
with $\omega_x/(2\pi) = 20$ Hz and $\omega_y/(2\pi) = \omega_z/(2\pi) = 400$ Hz. The
oscillator lengths with this set of parameters are $l_0 = 4.69~\mu$m and $l_{yz}= 1.05  ~\mu$m.
In numerical calculation it is found that all over  the ferromagnetic domain of Fig. \ref{fig-0} the density profiles are 
qualitatively the same always leading to a phase separation. In the anti-ferromagnetic domain the density profiles also remain similar without a phase separation. In the cyclic  domain the density profiles are different for $c_2>20c_1$  and $c_2<20c_1$,  the former leading to a phase separation and the latter leading to a  miscible configuration. We will consider these distinct  domains in the presentation 
of results.
In addition, in different domains we have different types of  
  symmetry-breaking  states, which we will  also illustrate.

The
background $s$-wave scattering lengths of $^{23}$Na in total spin $f_{\mathrm{tot}}  = 0,2,$ and $4$ channels
are $a_0=34.9a_B$, $a_2=45.8a_B$, and $a_4=64.5a_B$ \cite{ueda,Ciobanu}, respectively, where $a_B$ is the Bohr radius.
With these values of scattering lengths, we have $c_0 = 242.97, c_1 = 12.06 > 0,$ and $c_2 =-
13.03 < 20c_1$, corresponding to the anti-ferromagnetic domain in Fig. \ref{fig-0}. 
In the absence
of SO and Rabi  couplings, and magnetization, $\gamma = \Omega={\cal M} = 0$, there are more than one degenerate ground  states   \cite{ueda}. In Fig. \ref{fig-1} (a) - (c), three such degenerate
ground states are illustrated. 
These states are obtained with different initial choices for the 
wave-function  components in imaginary-time propagation. For these states, some 
of the wave-function  components have zero values.   
One can also have a state where the two components $j=\pm 1$ are populated with identical density 
(not illustrated here)
as in 
Fig. \ref{fig-1} (c), which can be generated by a  suitable rotation of 
the state of Fig. \ref{fig-1} (c) 
in spin space \cite{ueda}. 
In the presence of a non-zero SO coupling ($\gamma \ne 0, \Omega = {\cal M }=0$), 
	the degeneracy between the various ground states is removed 
 and only  the non-degenerate  state of 
Fig.~\ref{fig-1} (c) with the components $j=\pm 2$ survives.   
In this case there is no phase separation as concluded analytically  in Sec. 
\ref{Sec-IIB}. The degeneracy is also removed  for non-zero magnetization 
 ($\gamma =  \Omega = 0, {\cal M}\ne 0$). { In this case too,} 
there is only one ground state involving  components $j=\pm 2$, whose
density profile does not change with the introduction of SO coupling as is shown in Fig. \ref{fig-1} (d).
{ Hence,} the degenerate ground states exist only for zero magnetization in 
the absence of SO coupling.

\begin{figure}[!t]
\begin{center}
\includegraphics[trim = 5mm 0mm 4cm 0mm, clip,width=.49\linewidth,clip]{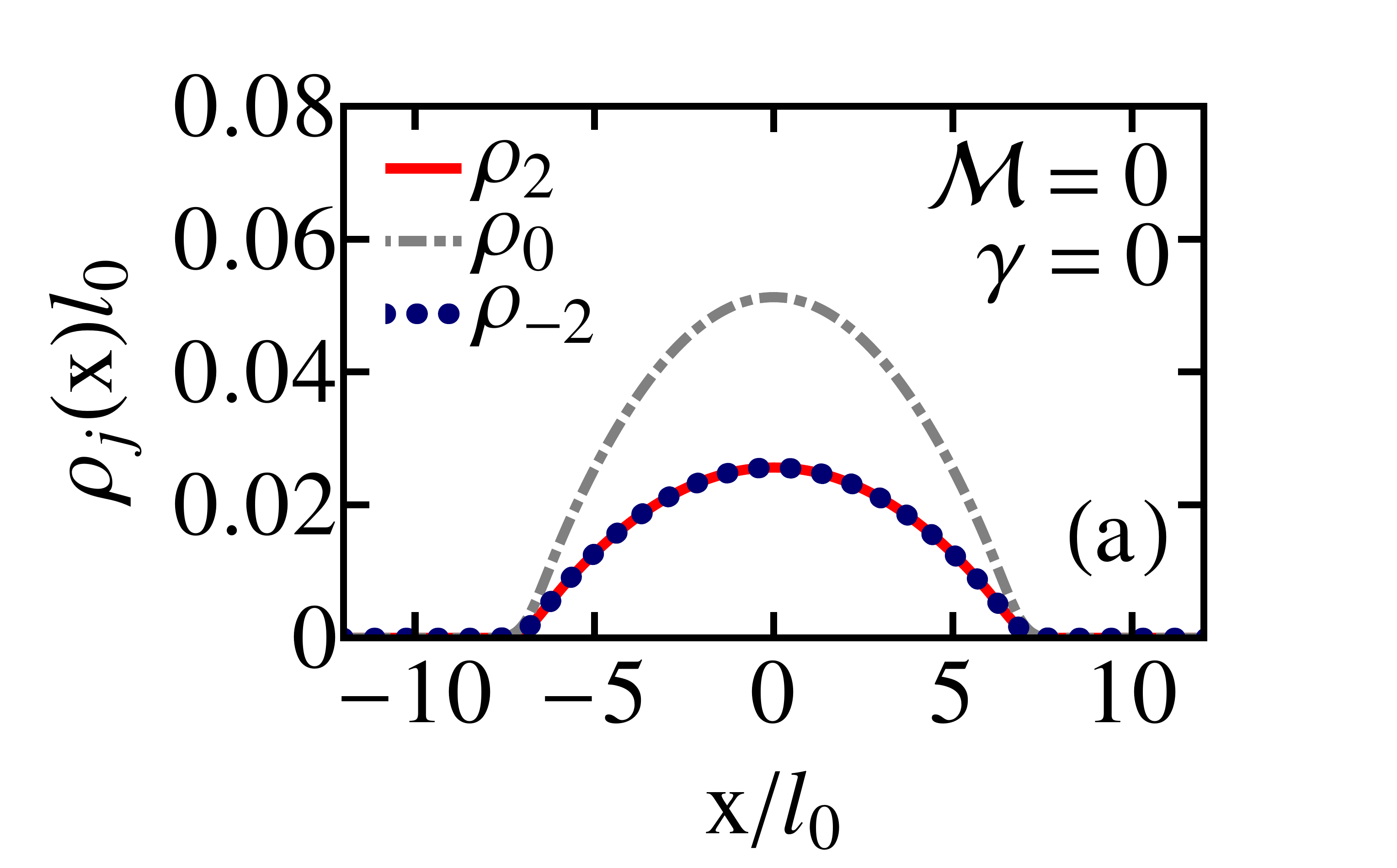}
\includegraphics[trim = 5mm 0mm 4cm 0mm, clip,width=.49\linewidth,clip]{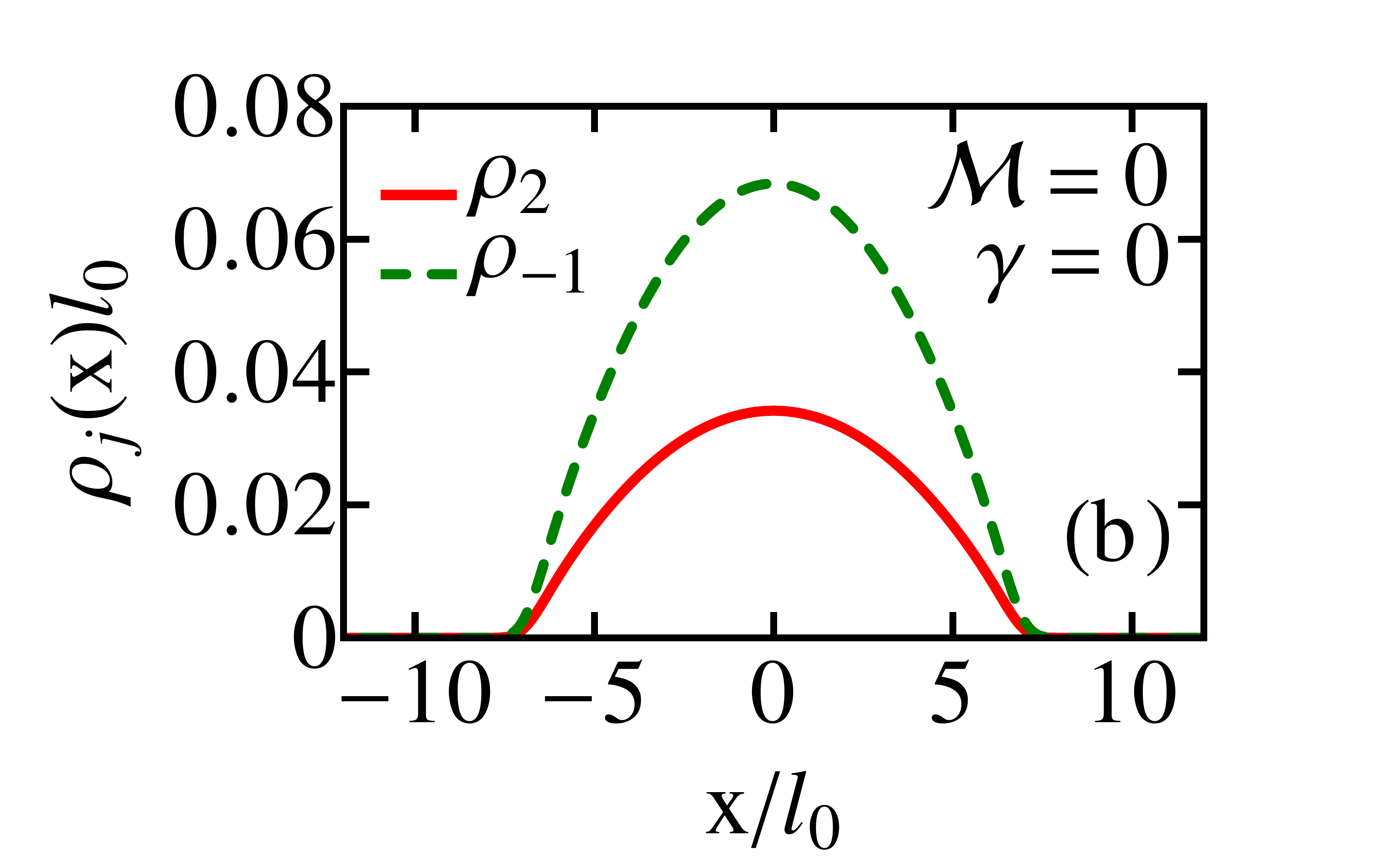}
\includegraphics[trim = 5mm 0mm 4cm 0mm, clip,width=.49\linewidth,clip]{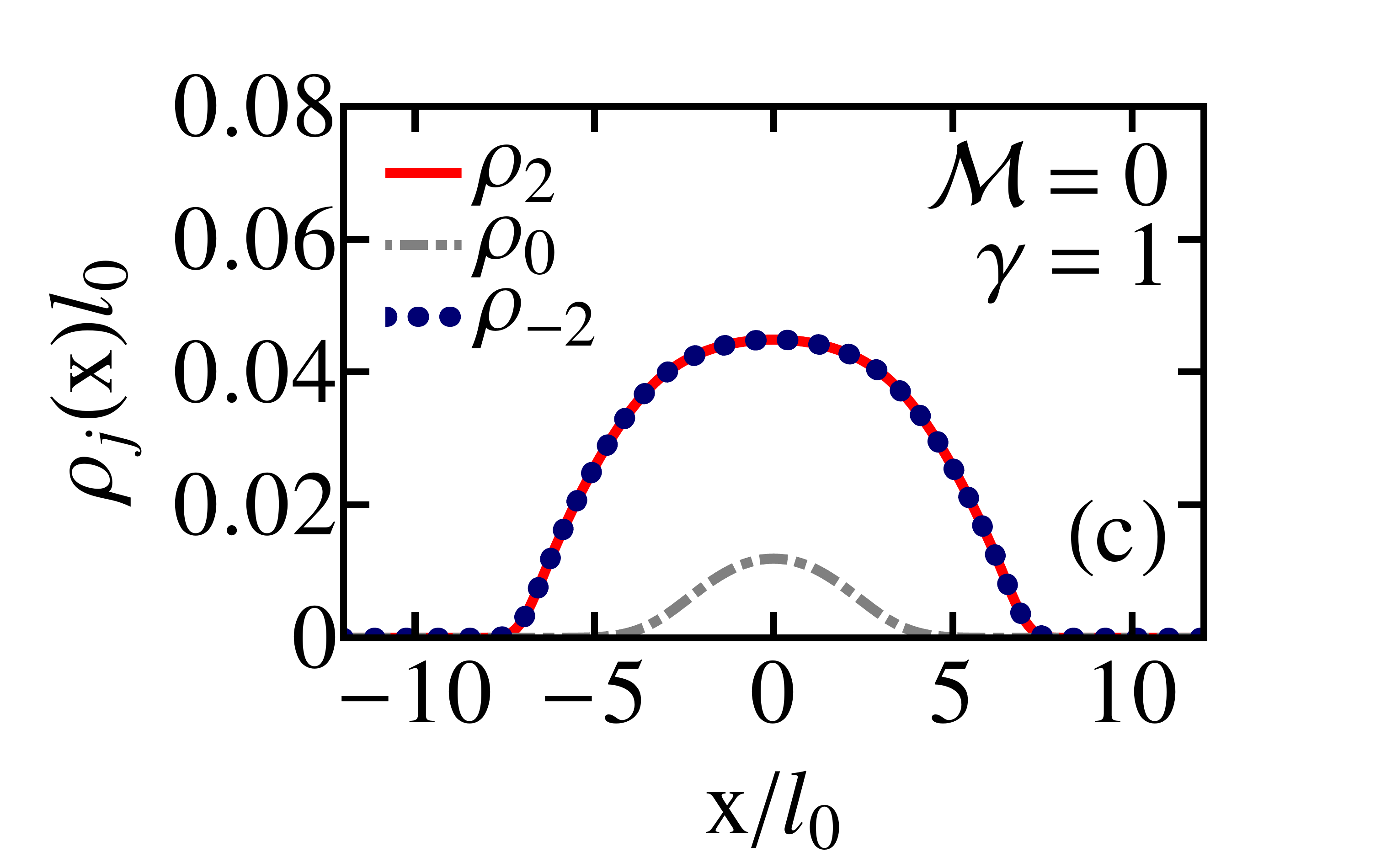}
\includegraphics[trim = 5mm 0mm 4cm 0mm, clip,width=.49\linewidth,clip]{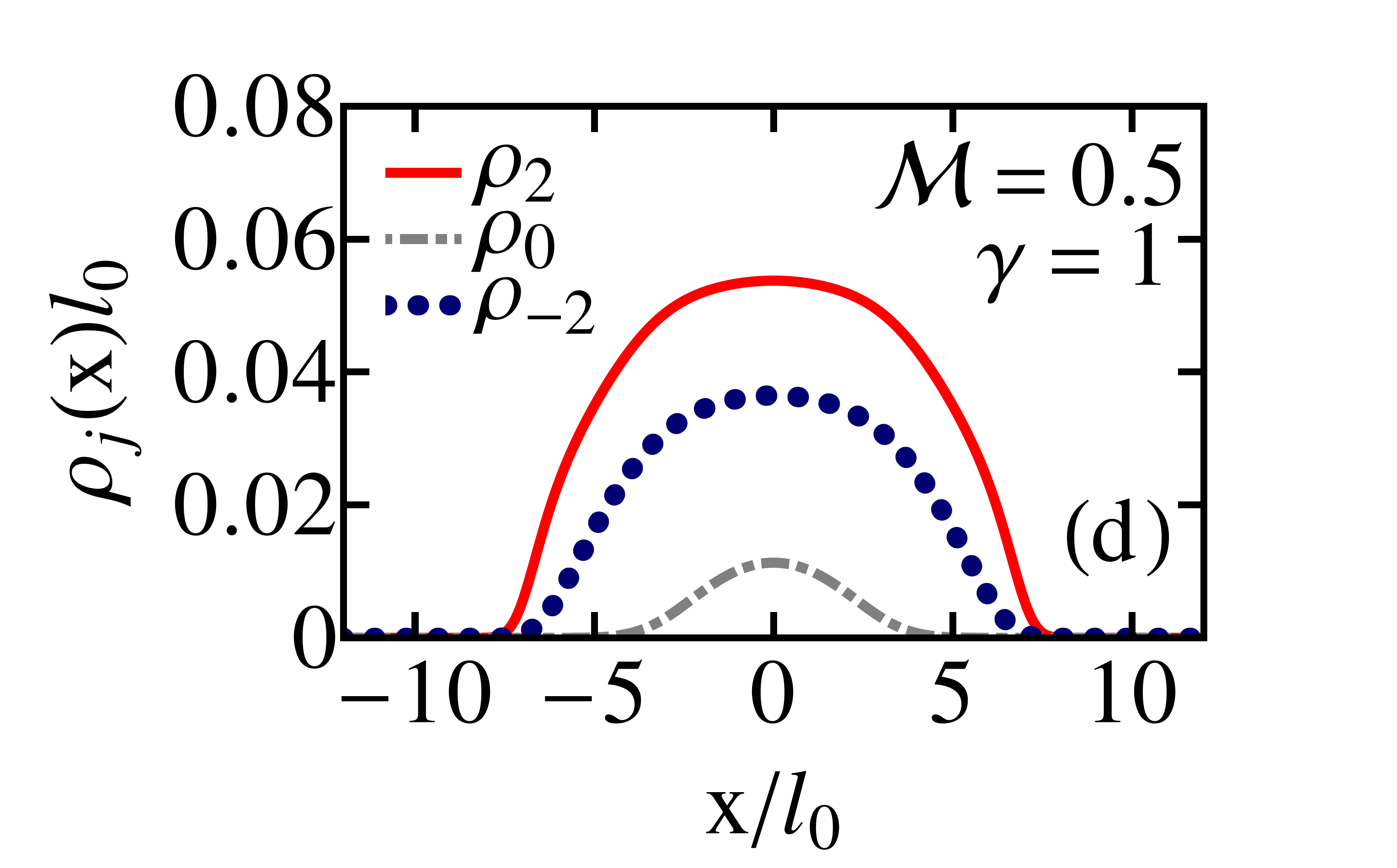} 
\caption{(Color online) Same as in Fig. \ref{fig-1} for  $ c_0=  242.97, c_1 = 12.06> 0$, and
$c_2 = 65.76<20c_1.$. The parameters used are 
(a) $\gamma={\cal M}=0$, (b)  $\gamma=0, {\cal M}=0$,     
(c)  $\gamma=1, {\cal M}=0$, and 
 (d)  $\gamma=1, {\cal M}=0.5$.    Time-reversal symmetry is broken in all cases.
 }
\label{fig-3} \end{center}
\end{figure}

To obtain $c_1>0$ and $0<c_2<20c_1$, we consider $a_0 = 52.35a_B$, 
$a_2 = 45.8a_B$, and $a_4 = 64.5a_B$, leading to $ c_0=  242.97, c_1 = 12.06> 0$, and
$c_2 = 65.76<20c_1.$ This corresponds to the cyclic domain with  miscible states. 
For ${\cal M} = 0$, the ground state solution in the absence of the SO coupling ($\gamma=0$) is shown in Fig. \ref{fig-3} (a),
which can be written as $(\phi_2,\phi_1,\phi_0,\phi_{-1},\phi_{-2})^T = \sqrt{\rho}(1,0,i\sqrt{2},0,1)^T/2$,
where $T$ stands for transpose and $\rho$ is the total density.
This state has  a purely imaginary 
$\phi_0$, a real $\phi_{\pm 2}$ and $\phi_{\pm 1}=0$,  and is 
degenerate with the state 
{  $(\phi_2,\phi_1,\phi_0,\phi_{-1},\phi_{-2})^T = \sqrt{\rho}(1,0,0,\sqrt{2},0)^T/\sqrt{3}$ shown} 
in Fig. \ref{fig-3} (b), where the $\pm j$ symmetry 
of the states is broken. The latter state has only $j=2,-1$ components.
{ The aforementioned two states break the time-reversal symmetry and are 
 degenerate} with their time-reversed counterparts.}  
{  For a nonzero $\gamma$, the degeneracy between these two states is no longer
ensured.}
With the introduction of a progressively increasing SO coupling $\gamma$, in the state of Fig. \ref{fig-3} (a)
$\rho_0$ decreases with
the corresponding increase in $\rho_{\pm 2}$  as is shown in Fig. \ref{fig-3} (c). With further increase in $\gamma$, $\rho_0$ becomes zero and only the components $j=\pm 2$ survive.
 The introduction of a nonzero magnetization ${\cal M}$ only introduces a splitting in the $j=\pm 2$ 
components as shown in Fig. \ref{fig-3} (d). 
There is no phase separation in this case also. Above a critical value of SO coupling $\gamma$, the condensate consists
of atoms in only $m_f=\pm 2$  states, { which for ${\cal M} = 0$ is time-reversal symmetric.}

\begin{figure}[!t]
\begin{center}
\includegraphics[trim = 5mm 0mm 4cm 0mm, clip,width=.49\linewidth,clip]{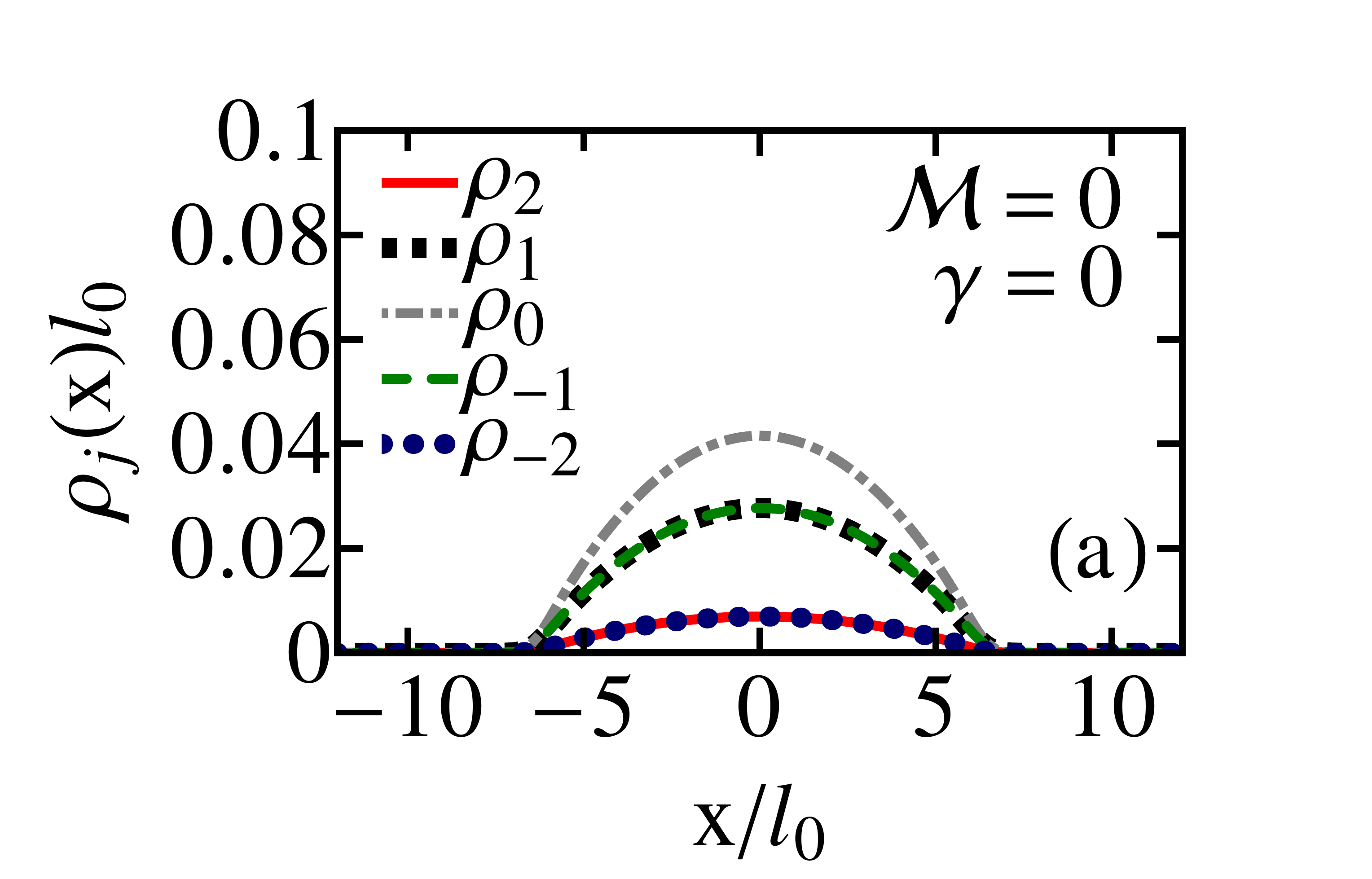}
\includegraphics[trim = 5mm 0mm 4cm 0mm, clip,width=.49\linewidth,clip]{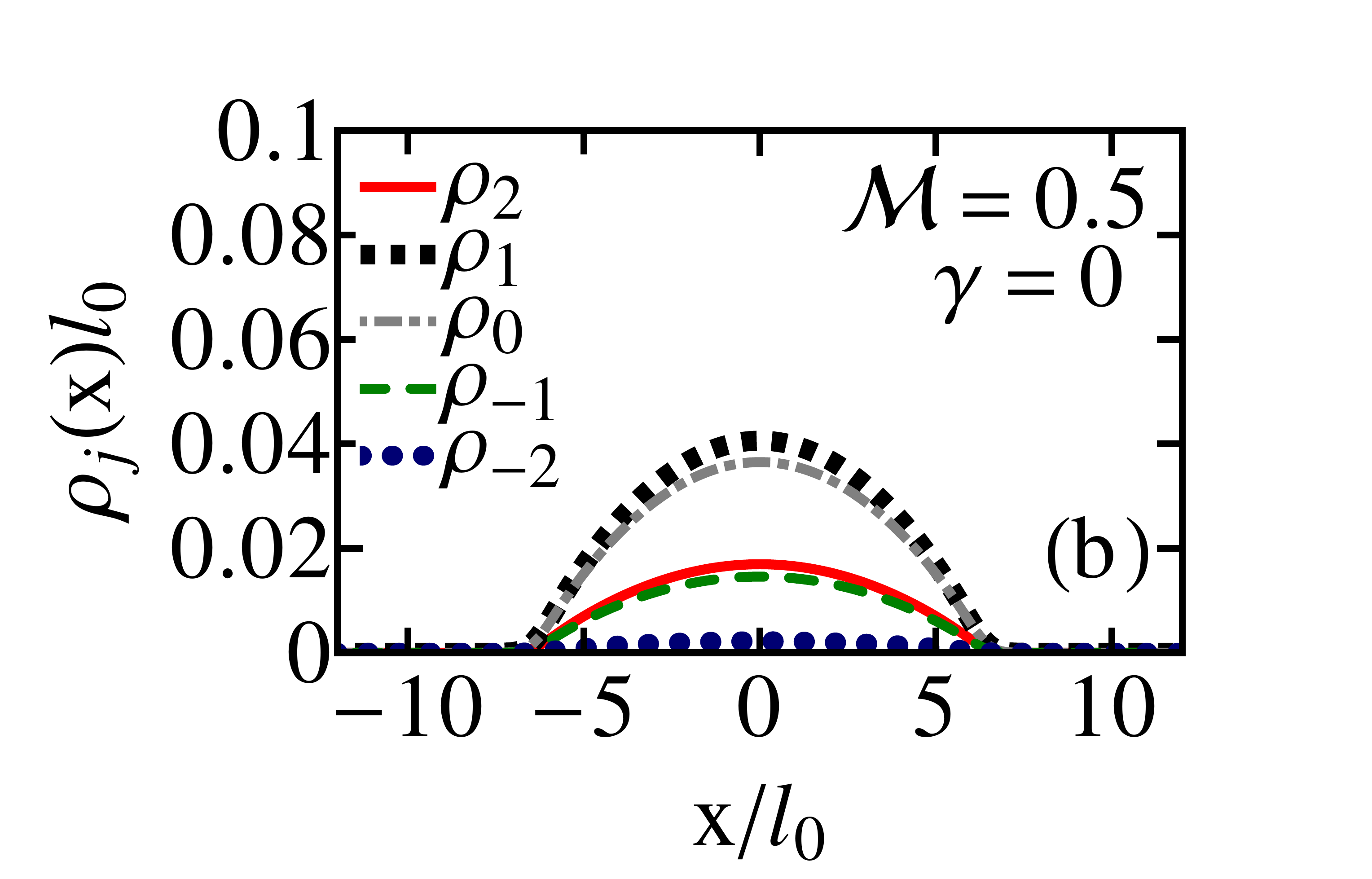}
\includegraphics[trim = 5mm 0mm 4cm 0mm, clip,width=.49\linewidth,clip]{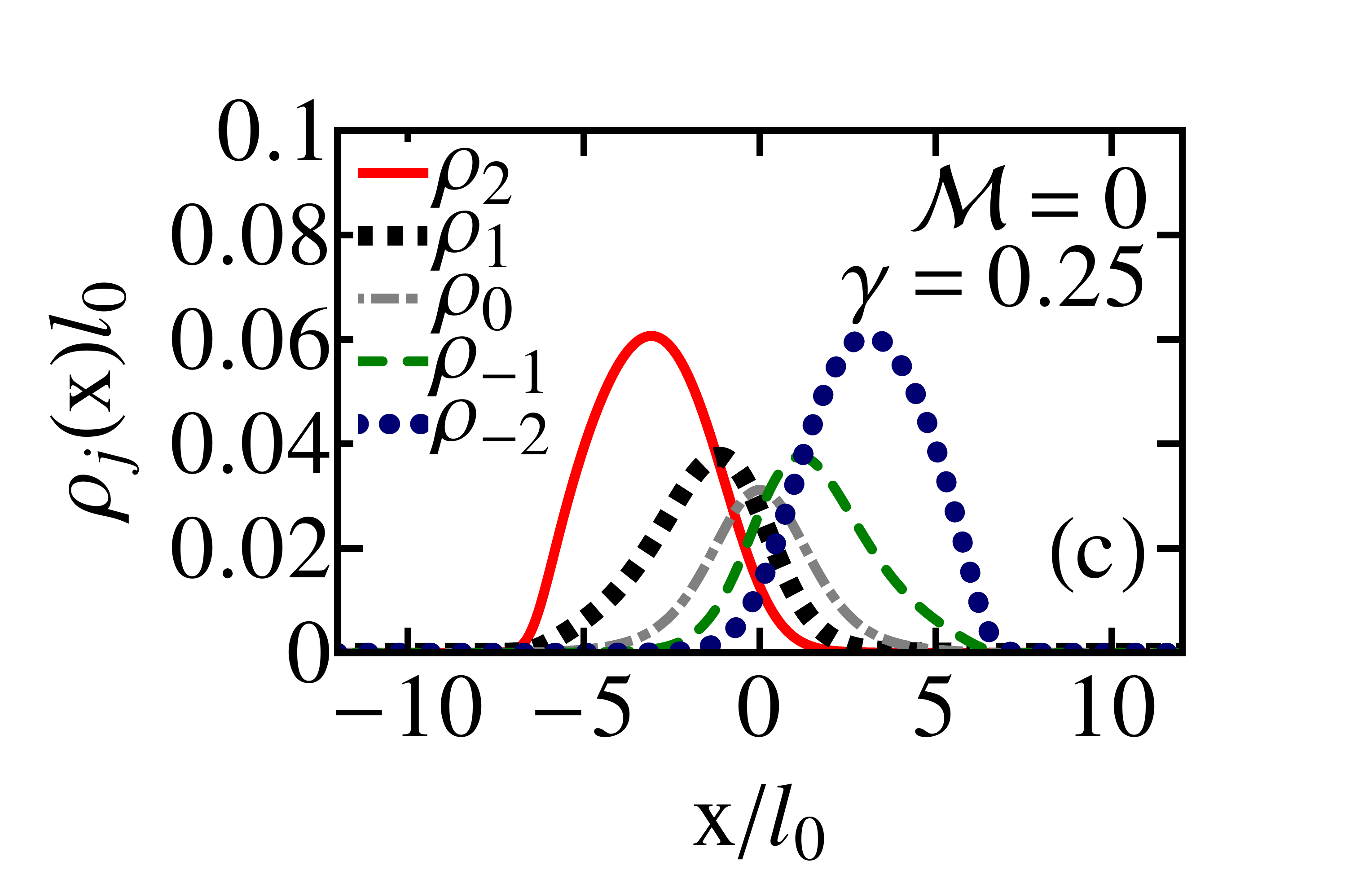}
\includegraphics[trim = 5mm 0mm 4cm 0mm, clip,width=.49\linewidth,clip]{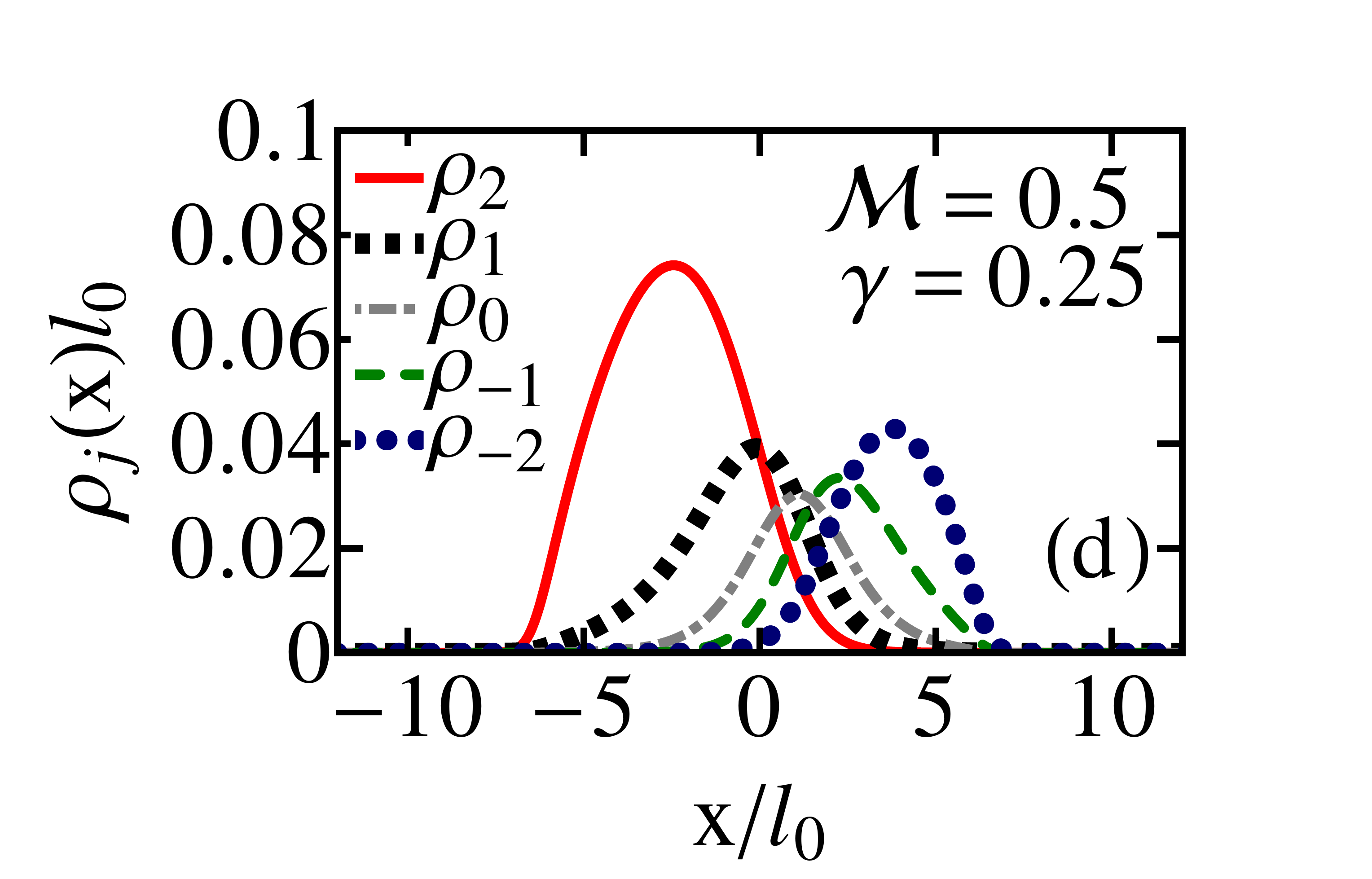}
\includegraphics[trim = 5mm 0mm 4cm 0mm, clip,width=.49\linewidth,clip]{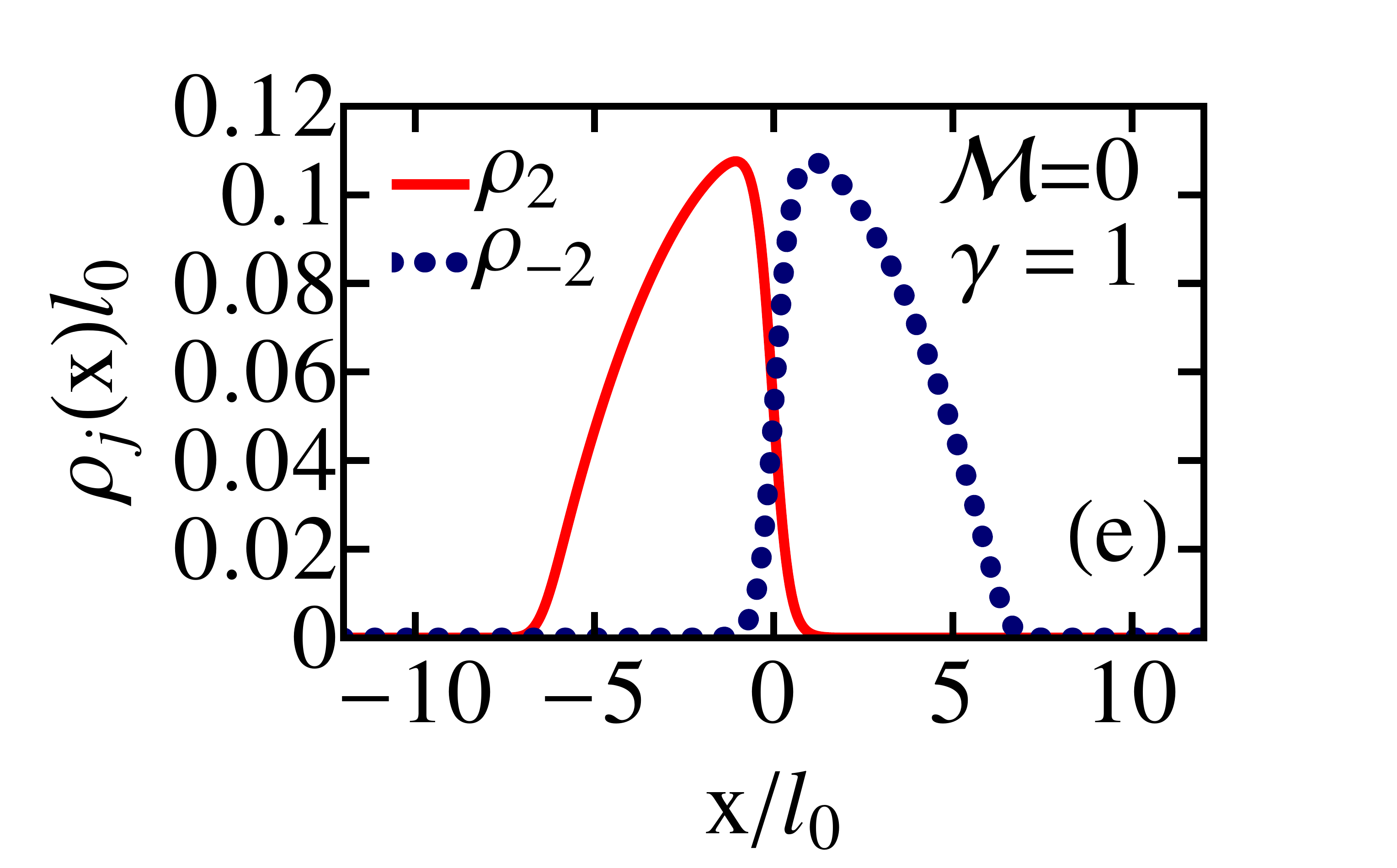}
\includegraphics[trim = 5mm 0mm 4cm 0mm, clip,width=.49\linewidth,clip]{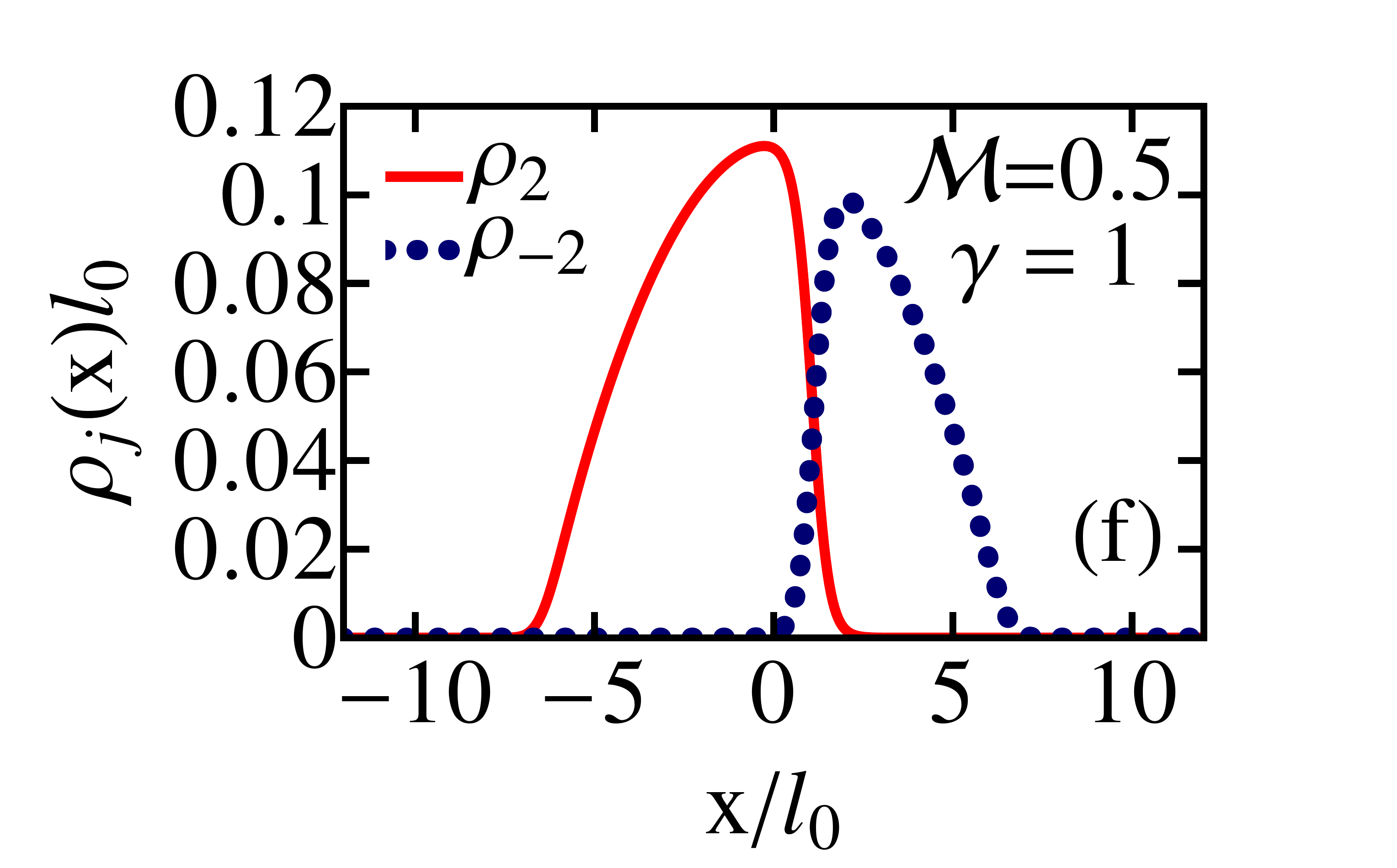}
\caption{(Color online) Same  as in Fig. \ref{fig-1} for $c_0= 201.36, c_1=-1.81,  c_2=24.15>20c_1$.
 The parameters used are 
(a) $\gamma={\cal M}=0$, (b)  $\gamma=0, {\cal M}=0.5$, (c)  $\gamma=0.25, {\cal M}=0$,
 (d)  $\gamma=0.25, {\cal M}=0.5$,    
(e)  $\gamma=1, {\cal M}=0$, and 
 (f)  $\gamma=1, {\cal M}=0.5$. Time-reversal symmetry is broken in (b) - (f).
 }
\label{fig-2} \end{center}
\end{figure}

The above  study shows that there cannot be a phase separation for $c_2<20c_1.$
Next  we  consider $c_2>20c_1.$ First we  consider a ferromagnetic state with 
 $c_1<0$ and $c_2>0$ obtained by employing  
 $a_0 = 52.35a_B$, $a_2 = 45.8a_B$, $a_4 = 43.0a_B$, leading to $c_0= 201.36, c_1=-1.81,  c_2=24.15>20c_1$.
In this case the densities of  the spinor BEC of 10000 atoms 
are shown for $\Omega =0$ for different values of  $\gamma  $ and ${\cal M}$.
in Fig. \ref{fig-2}   (a) - (f).  From these plots we see that with  an increase 
of SO coupling from $\gamma=0$, the overlapping 
component states separate and the population of the 
$j=\pm 2$ components increase at the cost of a reduction in population of the $j=0,\pm 1$ 
components. Finally, for $\gamma \approx > 1$ only the components  $j=\pm 2$ survive. In all cases 
a finite non-zero magnetization ${\cal M}$ breaks the symmetry between densities of the 
components $j=\pm 1$ and between $j=\pm 2$. 
The ground state solutions in this case are phase separated for $\gamma$ greater than a 
critical value in agreement with the discussion in Sec. \ref{Sec-IIB}.

\begin{figure}[!t]
\begin{center}
\includegraphics[trim = 5mm 0mm 4cm 0mm, clip,width=.48\linewidth,clip]{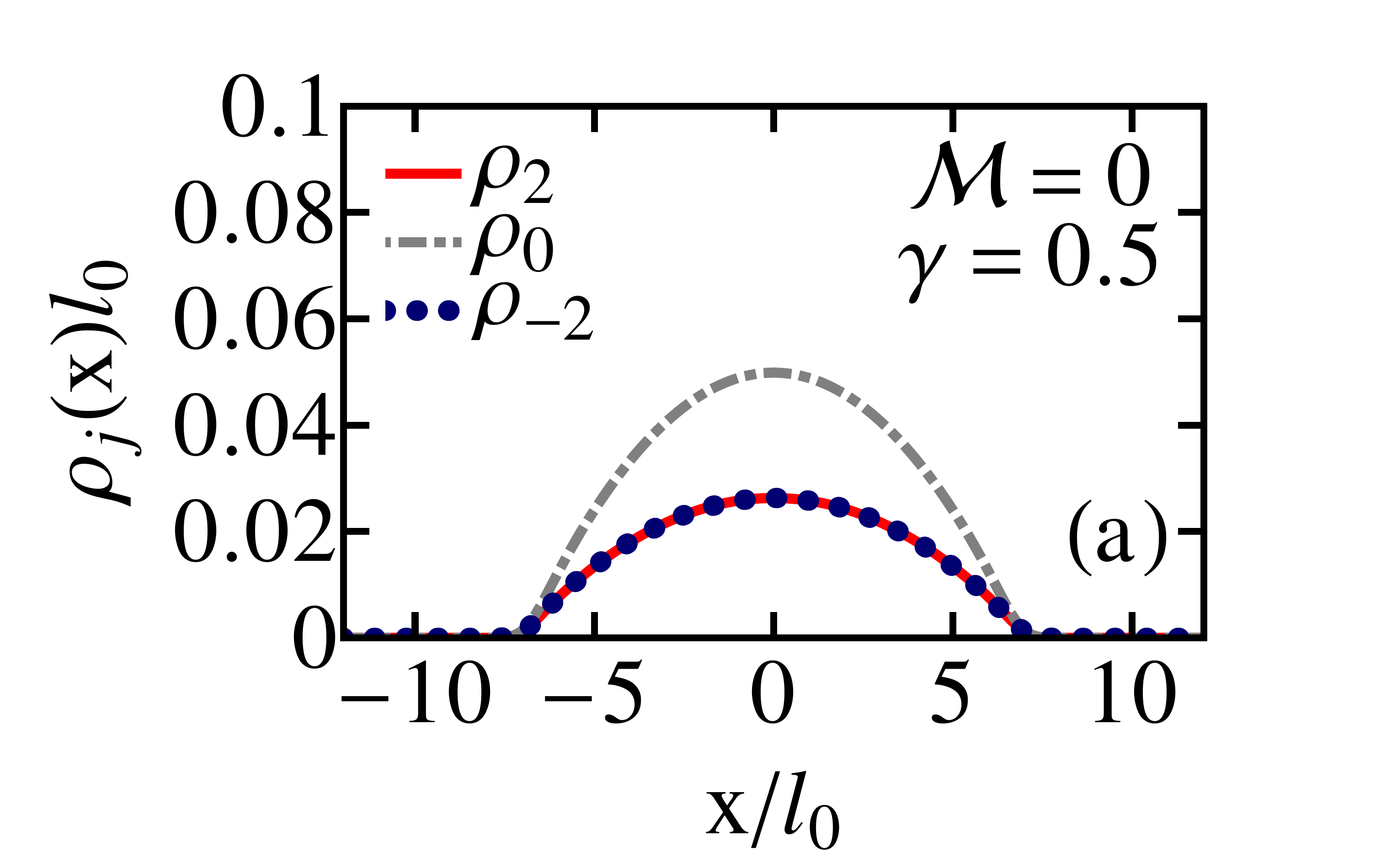}
\includegraphics[trim = 5mm 0mm 4cm 0mm, clip,width=.48\linewidth,clip]{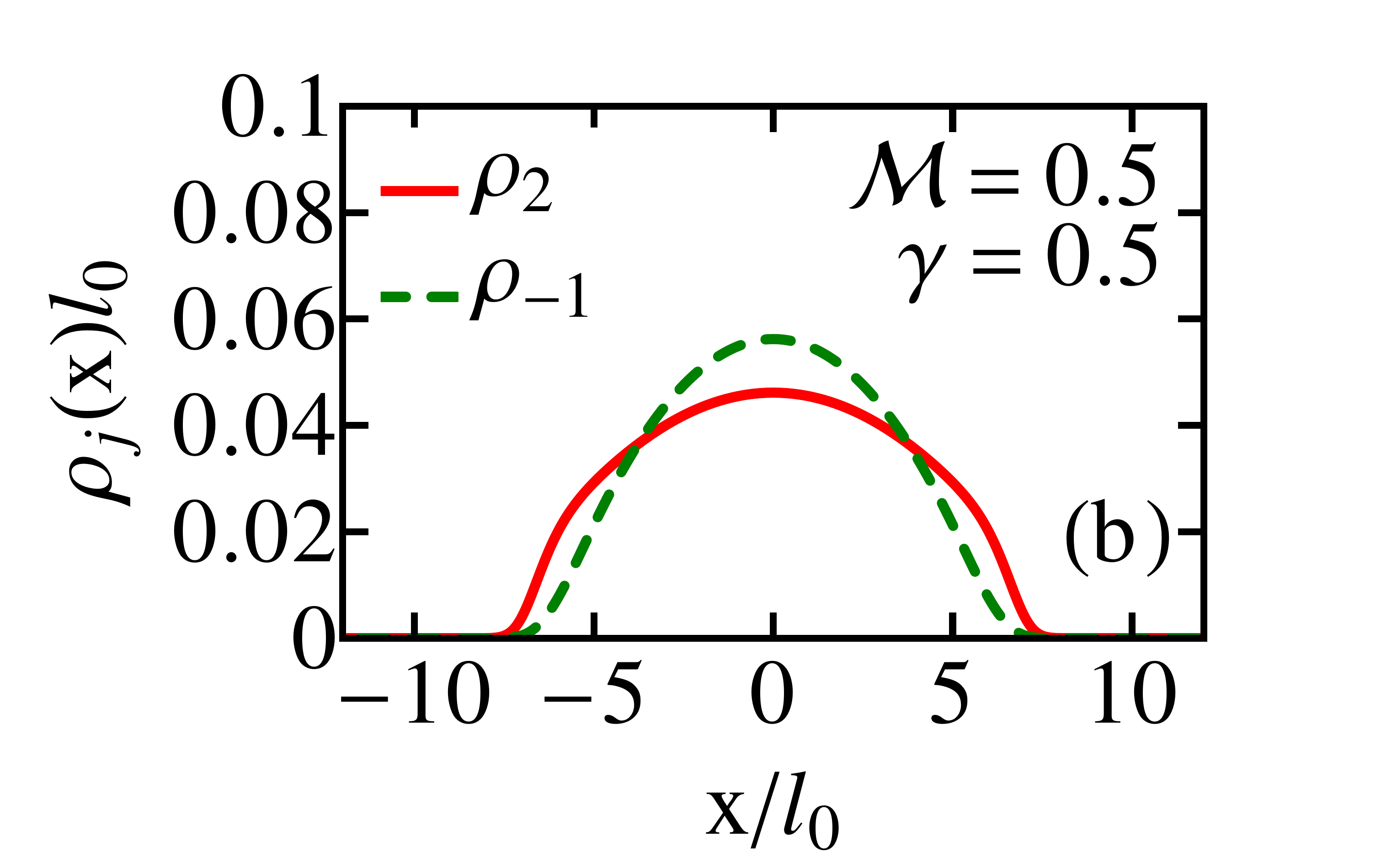}\\
\includegraphics[trim = 5mm 0mm 4cm 0mm, clip,width=.48\linewidth,clip]{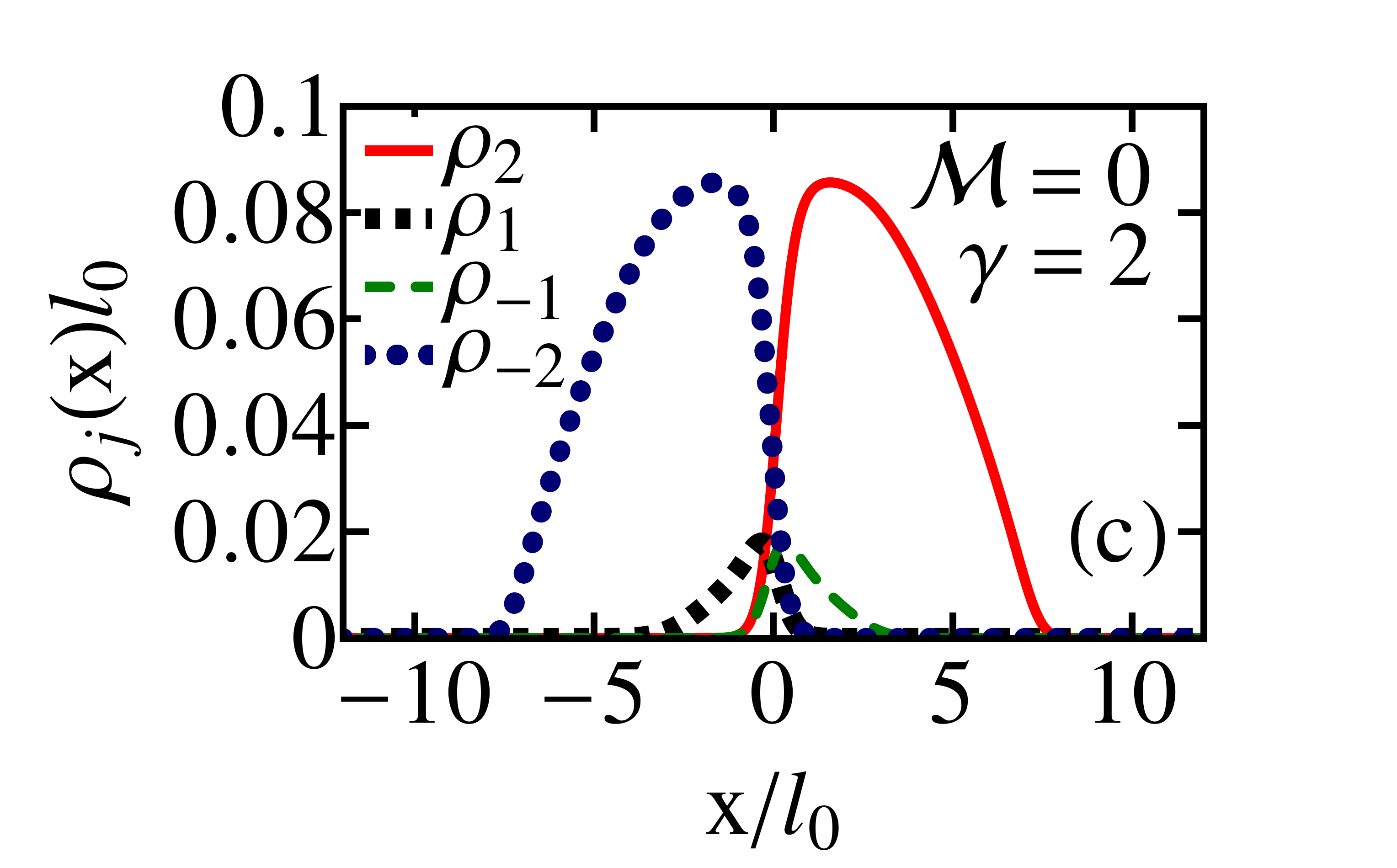}
\includegraphics[trim = 5mm 0mm 4cm 0mm, clip,width=.48\linewidth,clip]{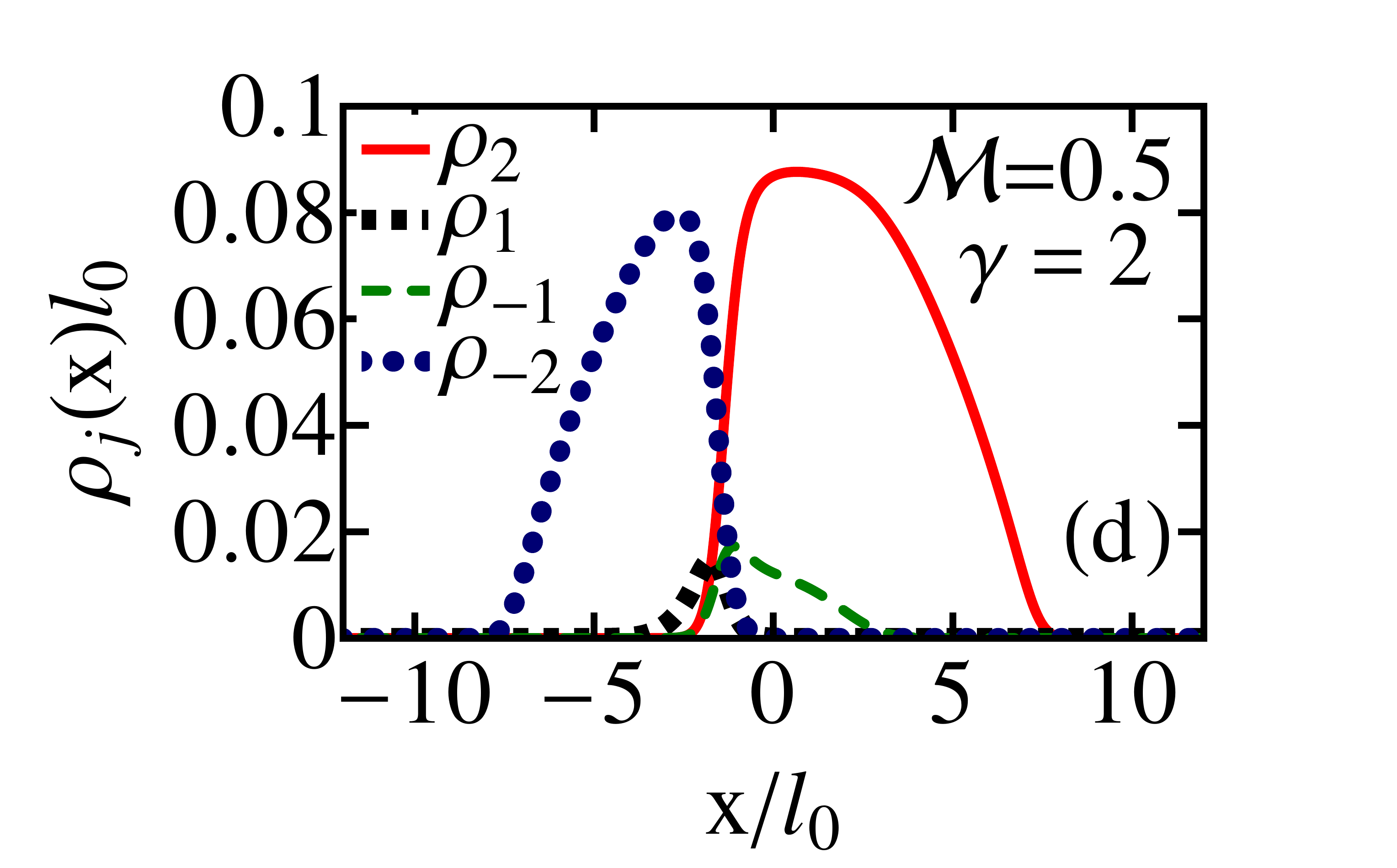}\\
\includegraphics[trim = 5mm 0mm 4cm 0mm, clip,width=.48\linewidth,clip]{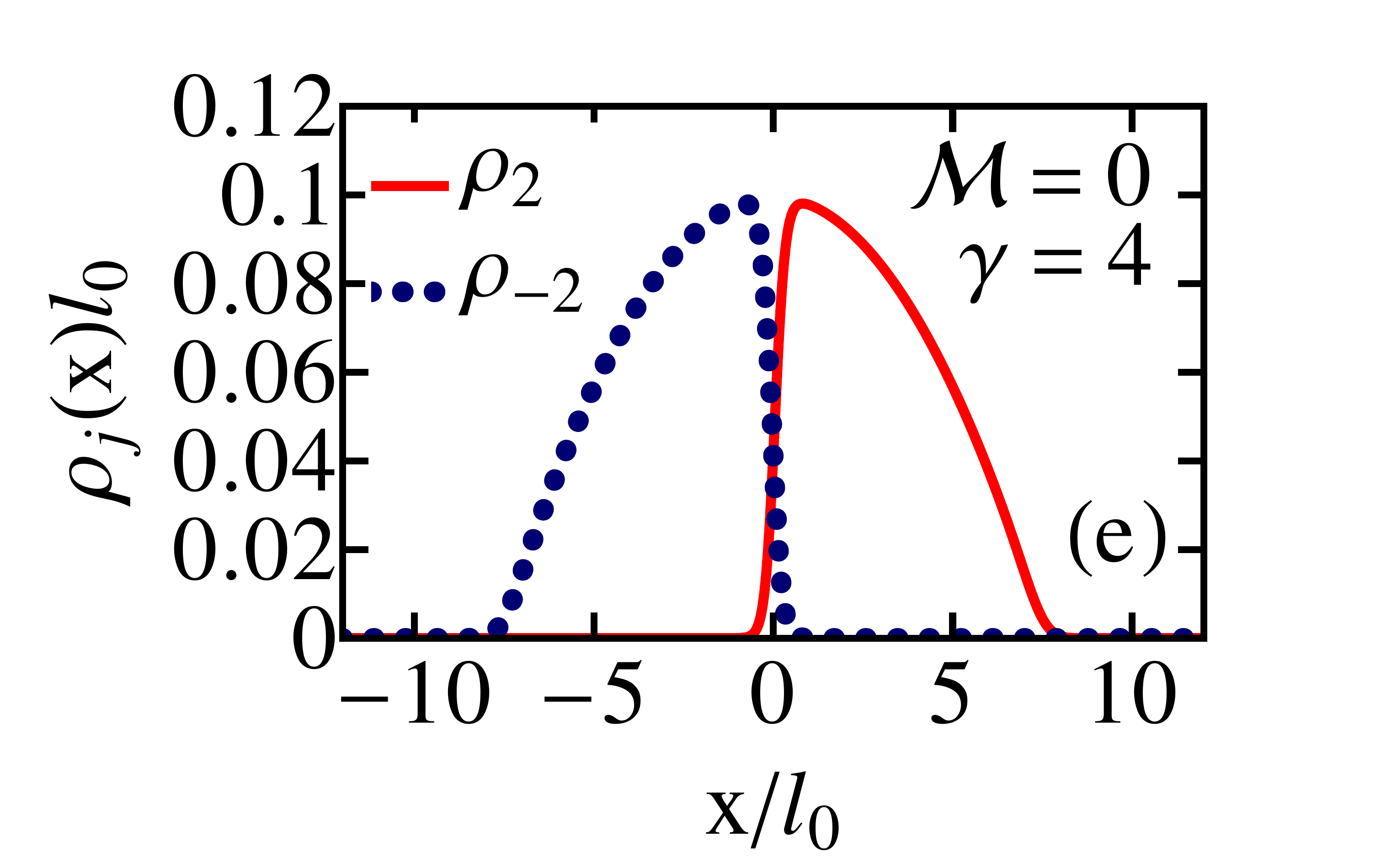}
\includegraphics[trim = 5mm 0mm 4cm 0mm, clip,width=.48\linewidth,clip]{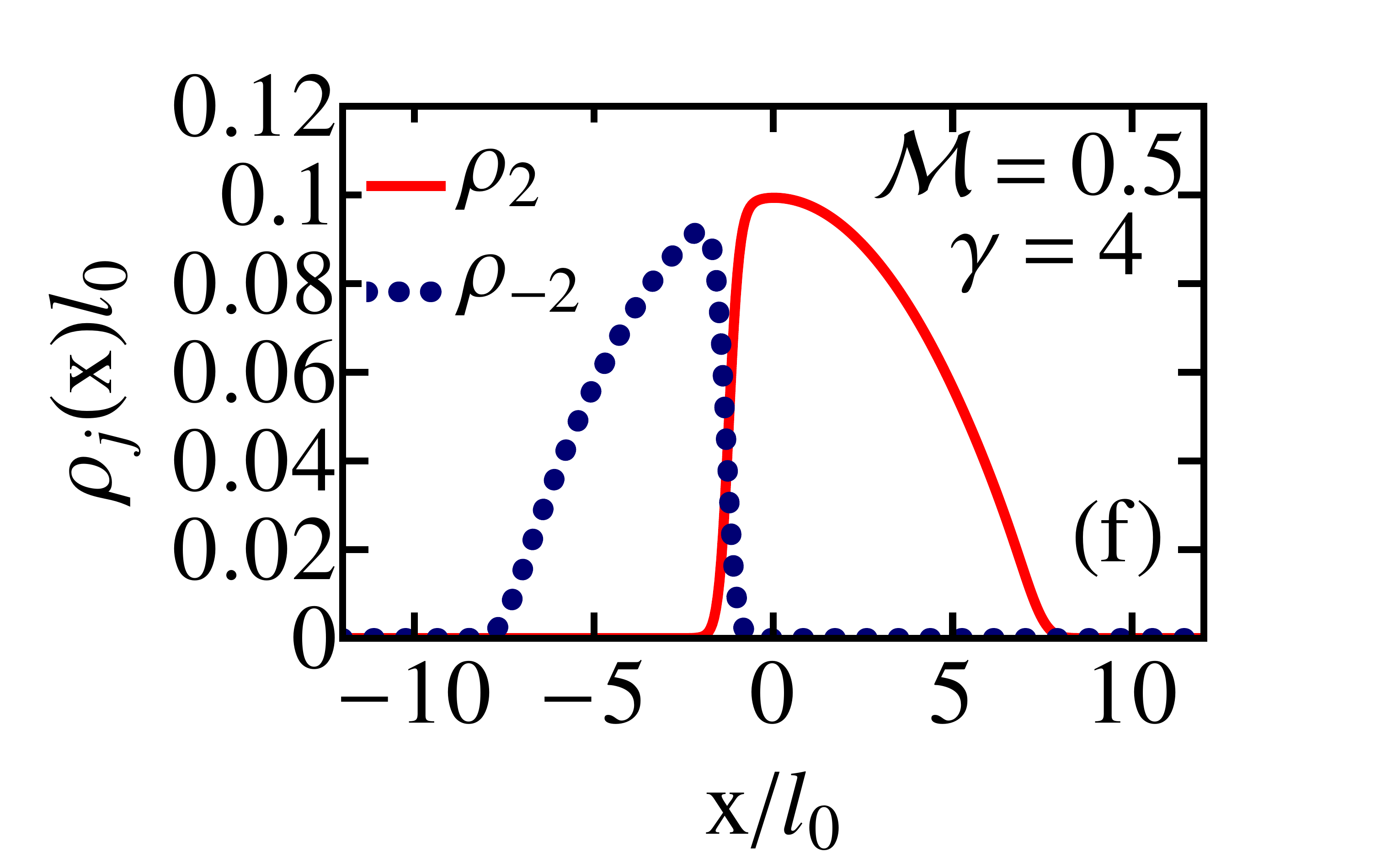}\\
\caption{(Color online) Same  as in Fig. \ref{fig-1} for $c_0=242.97>0, c_1=12.06>0$ and $c_2=459.68>20c_1$.
 The parameters used are 
(a) $\gamma=0.5, {\cal M}=0$, (b)  $\gamma=0.5, {\cal M}=0.5$, 
(c)  $\gamma=2, {\cal M}=0$,
 (d)  $\gamma=2, {\cal M}=0.5$,    
(e)  $\gamma=4, {\cal M}=0$, and 
 (f)  $\gamma=4, {\cal M}=0.5$.   Time-reversal symmetry is broken in all cases. }
\label{fig-4} \end{center}
\end{figure}

Next we consider a cyclic phase with $c_2>20c_1$. For this we consider $a_0 = 139.6a_0$, $a_2 = 45.8a_0$, and $a_4 = 64.5a_0$ leading to 
  $c_0=242.97>0, c_1=12.06>0$ and $c_2=459.68>20c_1$,  corresponding to a cyclic phase.
As the strength of SO coupling $\gamma$ is increased, there is a phase separation as is shown in Figs.
\ref{fig-4}
 (c) - (f). The nature of the
phase separation in this case is different from that discussed in $c_1<0$ and $c_2>20c_1$
in the sense that the components $j =2$ and $-1$ as well as $-2$ and $1$ are 
overlapping in this case, consistent with the conclusion of the theoretical analysis in Sec. \ref{Sec-IIB}.
{ The solutions in this domain always break time-reversal symmetry irrespective
of the value of $\gamma$.}

\begin{figure}[!t]
\begin{center}
\includegraphics[trim = 5mm 0mm 4cm 0mm, clip,width=.48\linewidth,clip]{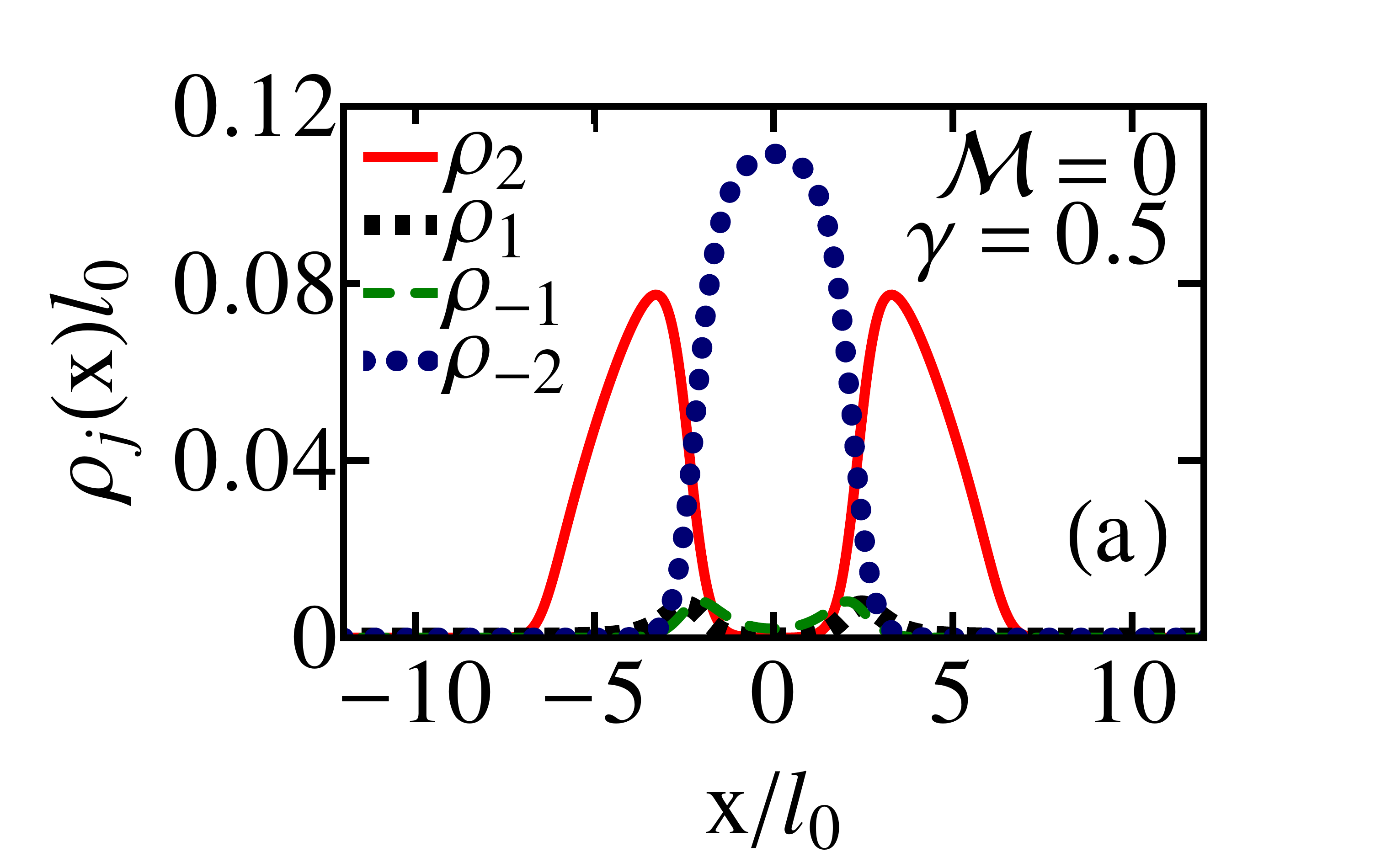}
\includegraphics[trim = 5mm 0mm 4cm 0mm, clip,width=.48\linewidth,clip]{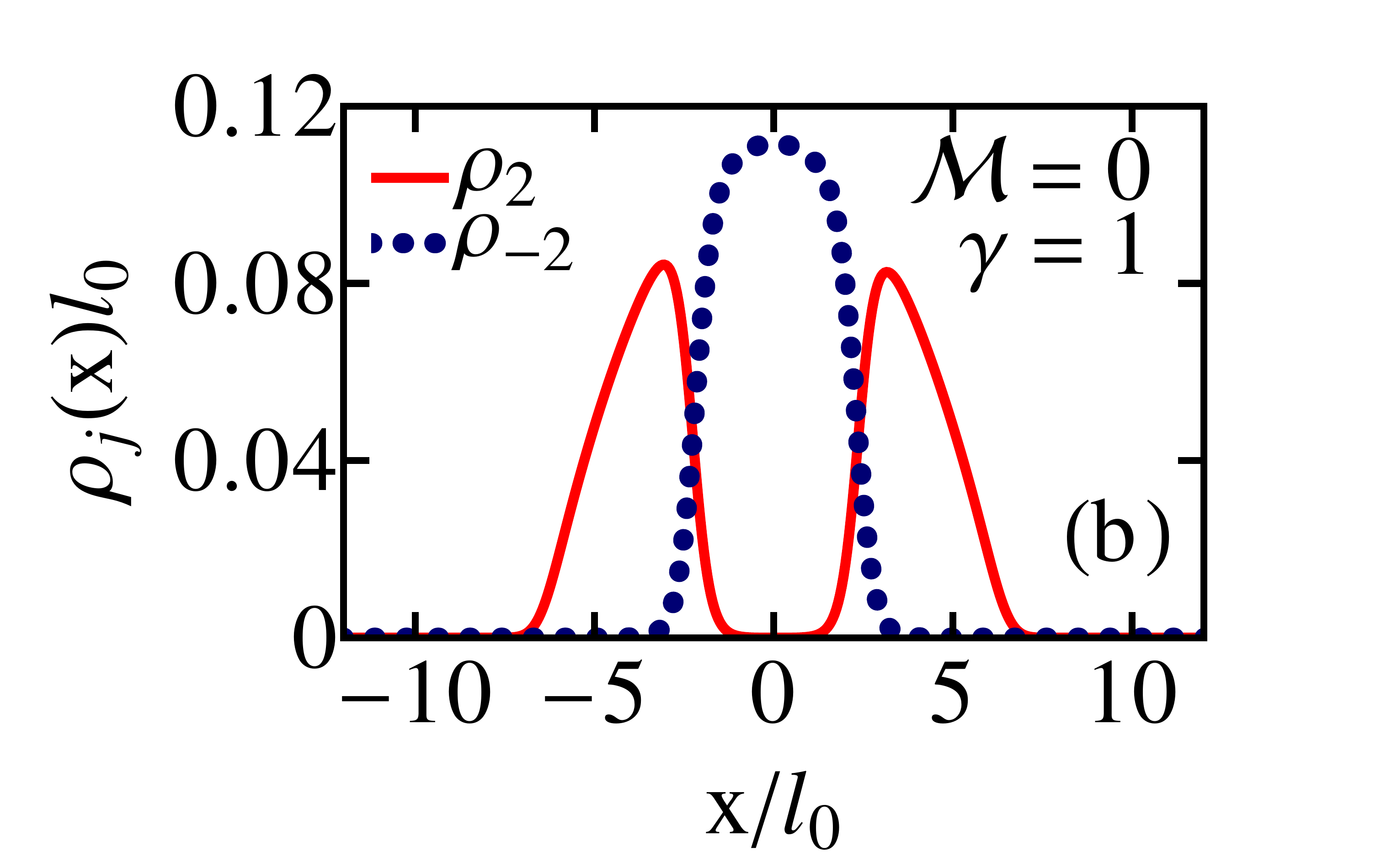}
\includegraphics[trim = 5mm 0mm 4cm 0mm, clip,width=.48\linewidth,clip]{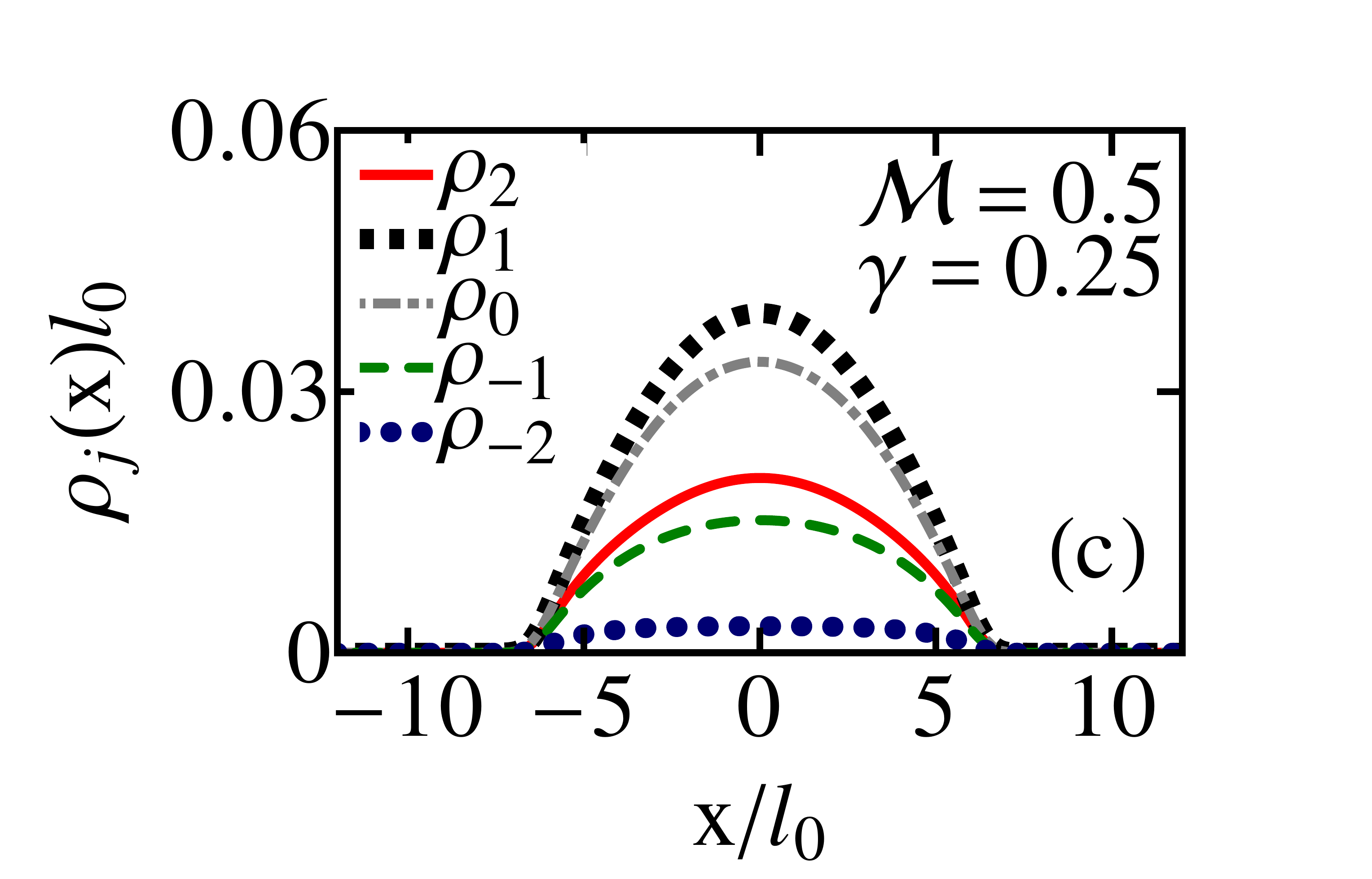}
\includegraphics[trim = 5mm 0mm 4cm 0mm, clip,width=.48\linewidth,clip]{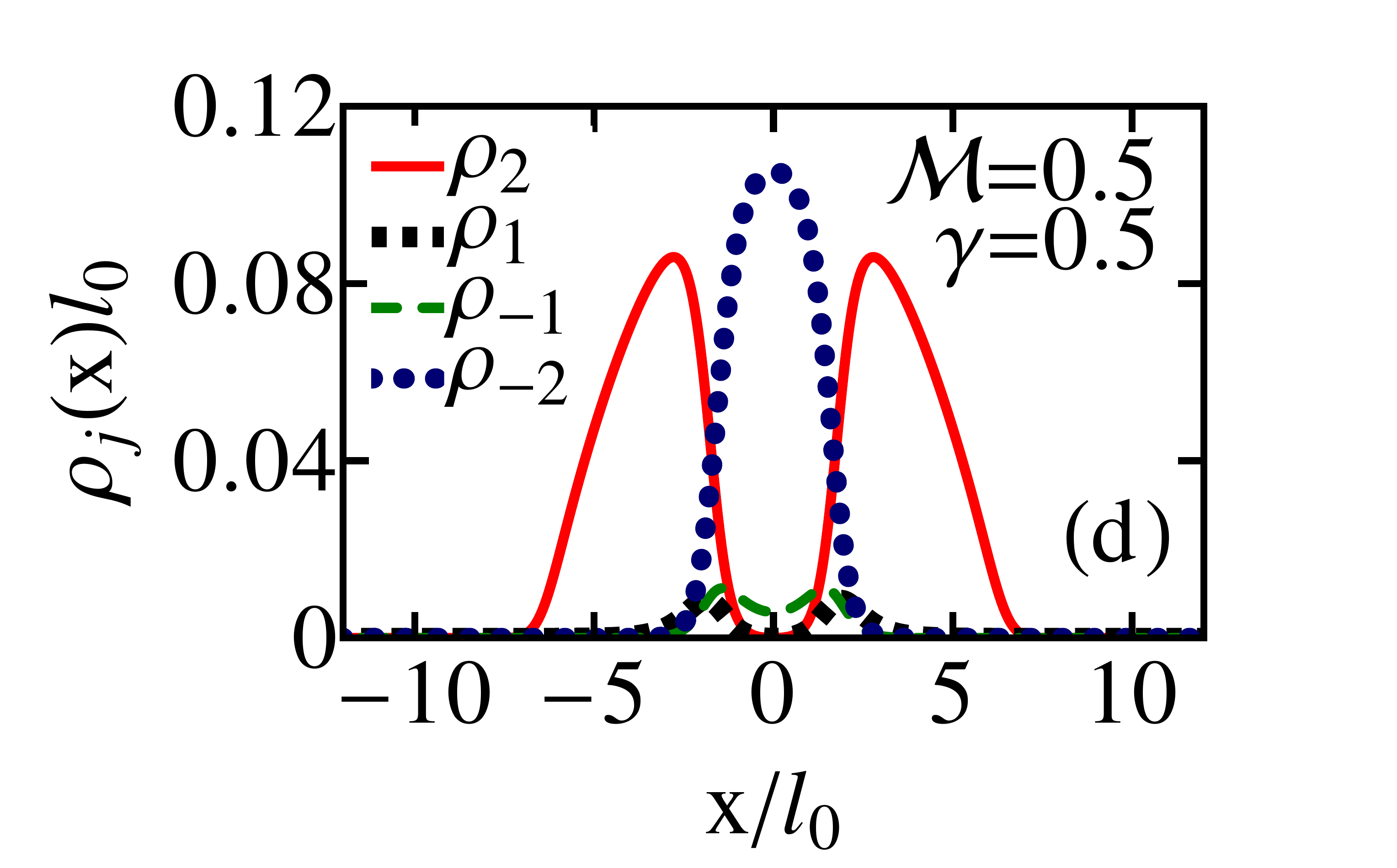}
\caption{(Color online)  Same  as in Fig. \ref{fig-1} for $c_0=201.36,c_1=-1.81,c_2=24.15>20c_1$.
 The parameters used are 
(a) $\gamma=0.5, {\cal M}=0$, (b)  $\gamma=1, {\cal M}=0$, 
(c)  $\gamma=0.25, {\cal M}=0.5$, and 
 (d)  $\gamma=0.5, {\cal M}=0.5$. Time-reversal symmetry is broken in all cases. 
  }
\label{fig-7} \end{center}
\end{figure}

{\em Symmetry-preserving versus symmetry-breaking solutions:} 
The imaginary-time propagation, that we use in calculation, preserves the symmetry of the initial input
wave function. Different  types of states can be obtained with different inputs.  For example, the states 
illustrated so far in Figs. \ref{fig-1} $-$ \ref{fig-4} were obtained with Gaussian inputs,
{ which were slightly shifted from the origin, for the component wave functions. 
We can obtain different density distribution for the components by using 
initial Gaussian inputs centered at the origin.} This is illustrated in Fig. \ref{fig-7} 
for the same set of parameters as in Fig. \ref{fig-2}: $c_0 = 201.36, c_1 = -1.81, c_2 = 24.15 > 20c_1$.
Distinct  from Fig.  \ref{fig-2}, in Fig.  \ref{fig-7}, the phase separation occurs in a different fashion preserving the symmetry of the trap: $\rho_j(x)=\rho_j(-x)$.  One of the components leaves the 
central region and stays symmetrically on both sides of the origin. The symmetry-breaking states of Fig.
  \ref{fig-2}  have lower energy than the symmetry-preserving states of Fig.  \ref{fig-7}
with the same sets of parameters. Similar symmetry-breaking states were found 
in scalar binary
condensates \cite{Esry}.
In symmetry-preserving and
symmetry-breaking cases the phase separation may start with different values of magnetization ${\cal M}$ and SO coupling $\gamma$.   The magnetization and SO couplings in Figs. \ref{fig-7} (b) and (c) are identical to those in  Figs. \ref{fig-2} (e) and (d), respectively. For the same set of parameters, 
there is a
phase separation in Fig.   \ref{fig-2} (d) and not in Fig. \ref{fig-7} (c). In Fig.  \ref{fig-7}
the phase separation starts at a larger value of SO coupling.

\begin{figure}[!t]
\begin{center}
\includegraphics[trim = 5mm 0mm 3cm 0mm, clip,width=.48\linewidth,clip]{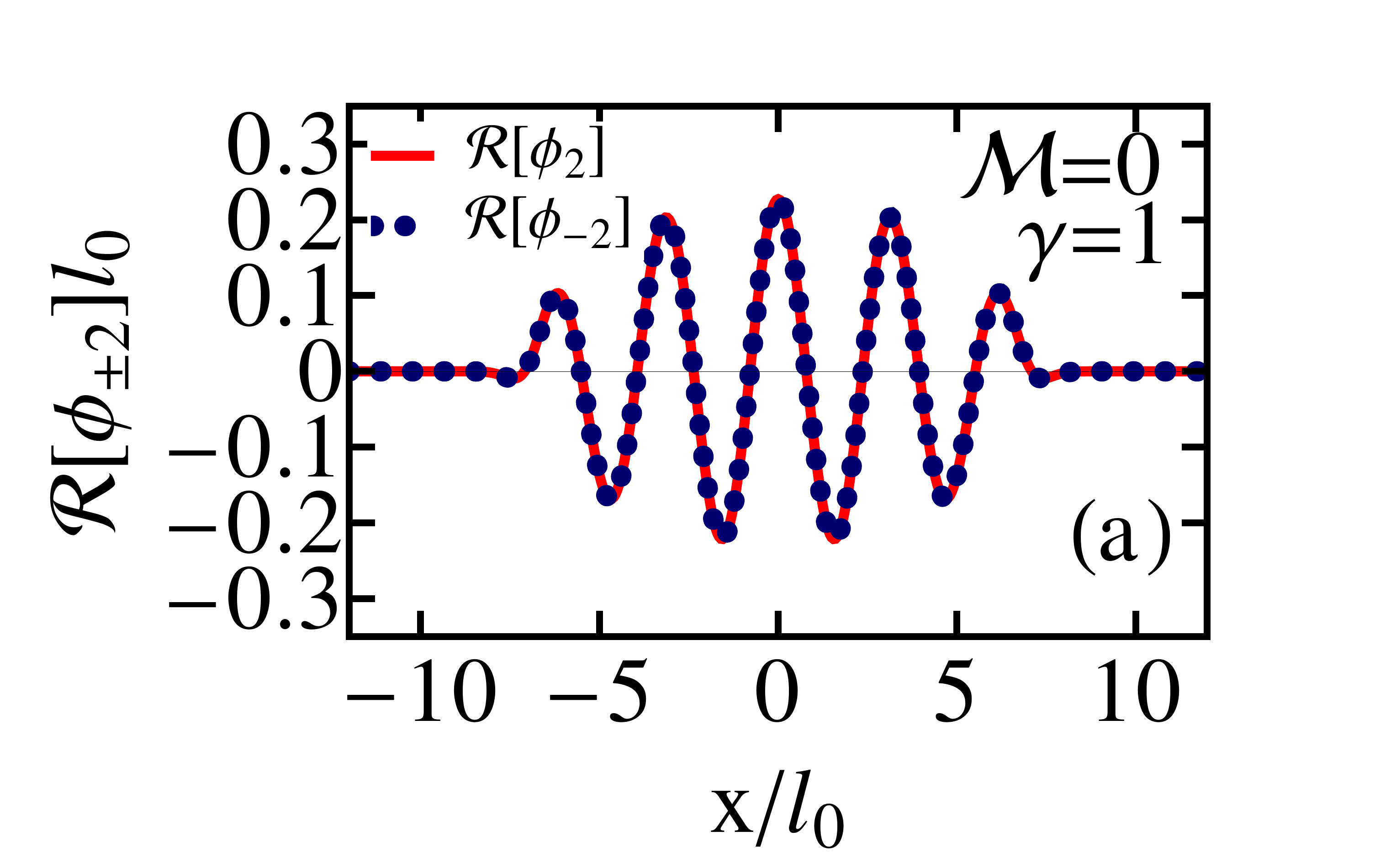}
\includegraphics[trim = 5mm 0mm 3cm 0mm, clip,width=.48\linewidth,clip]{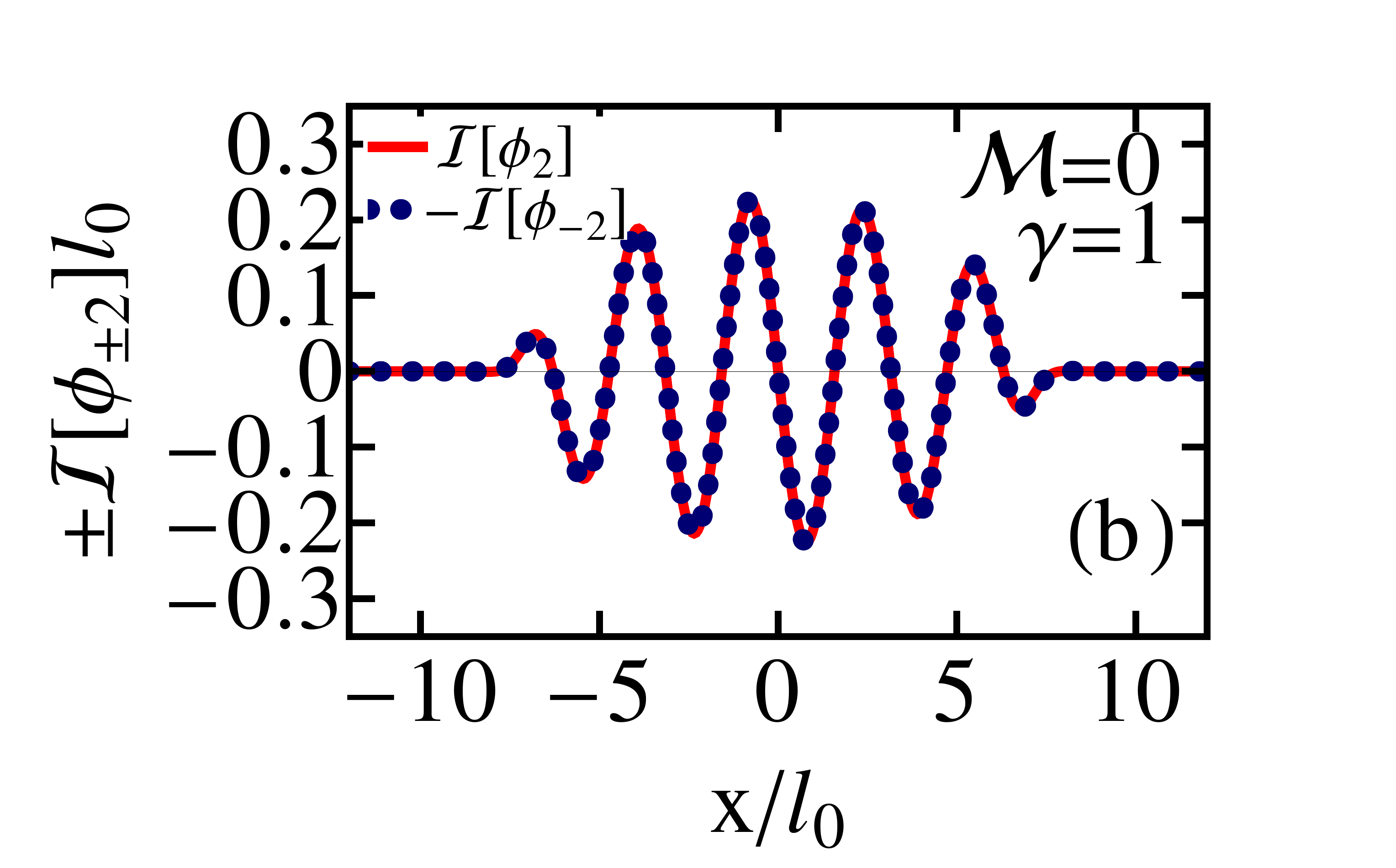}
\includegraphics[trim = 5mm 0mm 3cm 0mm, clip,width=.48\linewidth,clip]{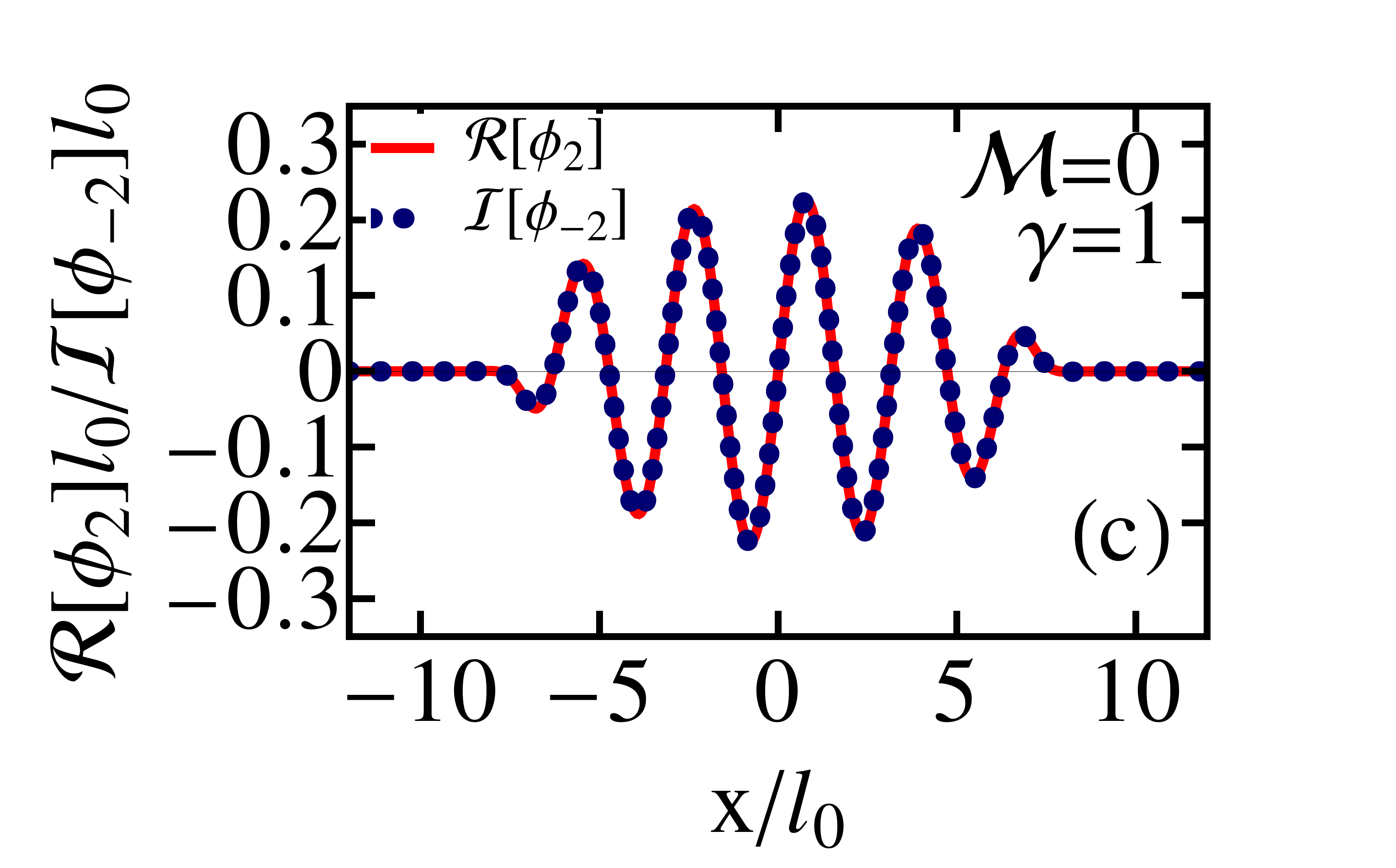}
\includegraphics[trim = 5mm 0mm 3cm 0mm, clip,width=.48\linewidth,clip]{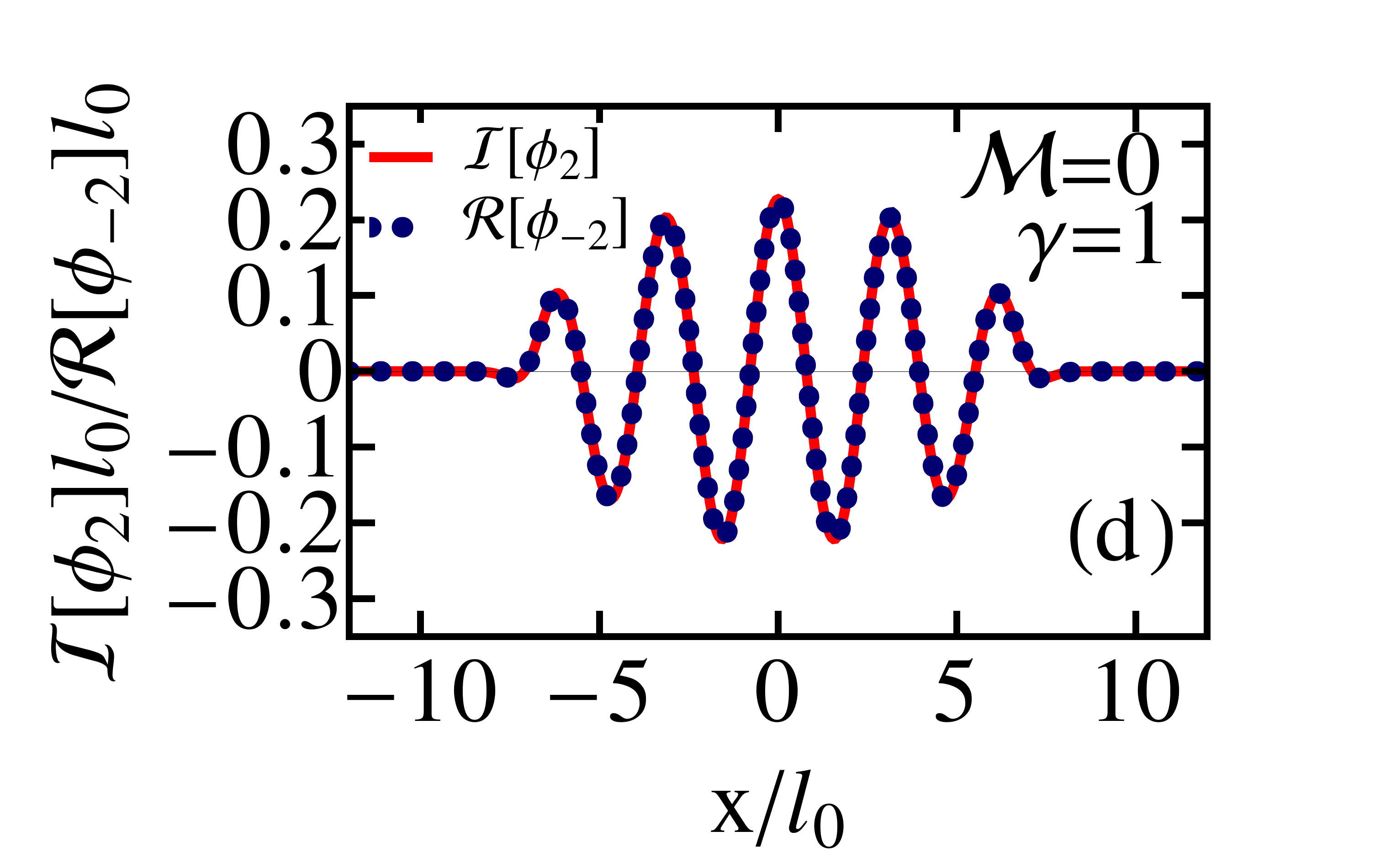}
\caption{ Real and imaginary parts of the $j=\pm 2$  wave-function components 
for the parameters of Fig. \ref{fig-1} (c). Plots  (a) and (b) were obtained with 
two real Gaussian input states for $j=\pm 2$ components. Plots (c) and (d) were obtained with 
an imaginary Gaussian input for $j=2$ and a real Gaussian input for $j=-2$ components.
}
\label{fig-9} \end{center}
\end{figure}

Next we show in Fig. \ref{fig-9} some of the possible parity-breaking states obtained with the parameters of Fig. 
\ref{fig-1} (c), e.g., $c_0=242.97, c_1=12.06>0, c_2=-13.03<20c_1$. Figures \ref{fig-9} (a) and (b) were 
calculated with real Gaussian inputs for both the $j=\pm 2$ components.  This will correspond to 
$\alpha_2 =\alpha_{-2}=1/\sqrt 2$ in the discussion in Sec. \ref{Sec-IIB}, leading to symmetry 
properties  ${\cal R} [\phi_{\pm 2}(x)]=
  {\cal R} [\phi_{\pm 2}(-x)], $ $
 {\cal I} [\phi_{\pm 2}(x)]=-
  {\cal I} [\phi_{\pm 2}(-x)],$ $
 {\cal R} [\phi_{2}(x)]=
  {\cal R} [\phi_{- 2}(x)], $ $
 {\cal I} [\phi_{ 2}(x)]=-
  {\cal I} [\phi_{- 2}(x)]$, as illustrated in Fig. \ref{fig-9} (a) and (b).
Figures \ref{fig-9} (c) and (d) were 
calculated with an imaginary Gaussian input for $j=2$ and a real Gaussian input for   $j=- 2$ components.
  This will correspond to $\alpha_2 =i/\sqrt 2 $ and $   \alpha_{-2}=1/\sqrt 2$ in the discussion in 
Sec. \ref{Sec-IIB}, leading to symmetry properties
${\cal R} [\phi_{ 2}(x)]=
-  {\cal R} [\phi_{ 2}(-x)], $ 
${\cal R} [\phi_{ -2}(x)]=
  {\cal R} [\phi_{- 2}(-x)], $ 
${\cal I} [\phi_{ 2}(x)]=
  {\cal I} [\phi_{ 2}(-x)], $ 
${\cal I} [\phi_{ -2}(x)]=
  -{\cal I} [\phi_{- 2}(-x)], $ 
$
 {\cal R} [\phi_{2}(x)]=
  {\cal I} [\phi_{-2}(x)],$ $
 {\cal I} [\phi_{2}(x)]=  
  {\cal R} [\phi_{- 2}(x)], $    
as illustrated in Fig. \ref{fig-9} (c) and (d).
In the same fashion all the parity-breaking states obtained  in Sec. \ref{Sec-IIB} from an analytic consideration 
can be realized numerically.

As in the case of a binary condensate \cite{Ao}, phase-separated SO-coupled
spinor condensates can be categorized either as weakly segregated or strongly
segregated. In the weakly segregated domain, the total density profile preserves
the symmetry of the trapping potential (not illustrated here)
and has an approximate smooth Gaussian profile
like that of a single-component BEC in a trap.  
In the strongly segregated domain, a notch appears
in the total density profile corresponding to symmetry breaking solutions
shown in Figs. \ref{fig-2} (e), (f), \ref{fig-4} (e) and \ref{fig-4} (f). 
In case of a non-zero magnetization, asymmetrically 
located notch ensures that the total density profiles corresponding
to Figs. \ref{fig-2} (f) and \ref{fig-4} (f) do not not have the symmetry 
of the trap (not  shown here).

\begin{figure}[!t]
\begin{center}
\includegraphics[trim = 5mm 0mm 4cm 0mm, clip,width=.48\linewidth,clip]{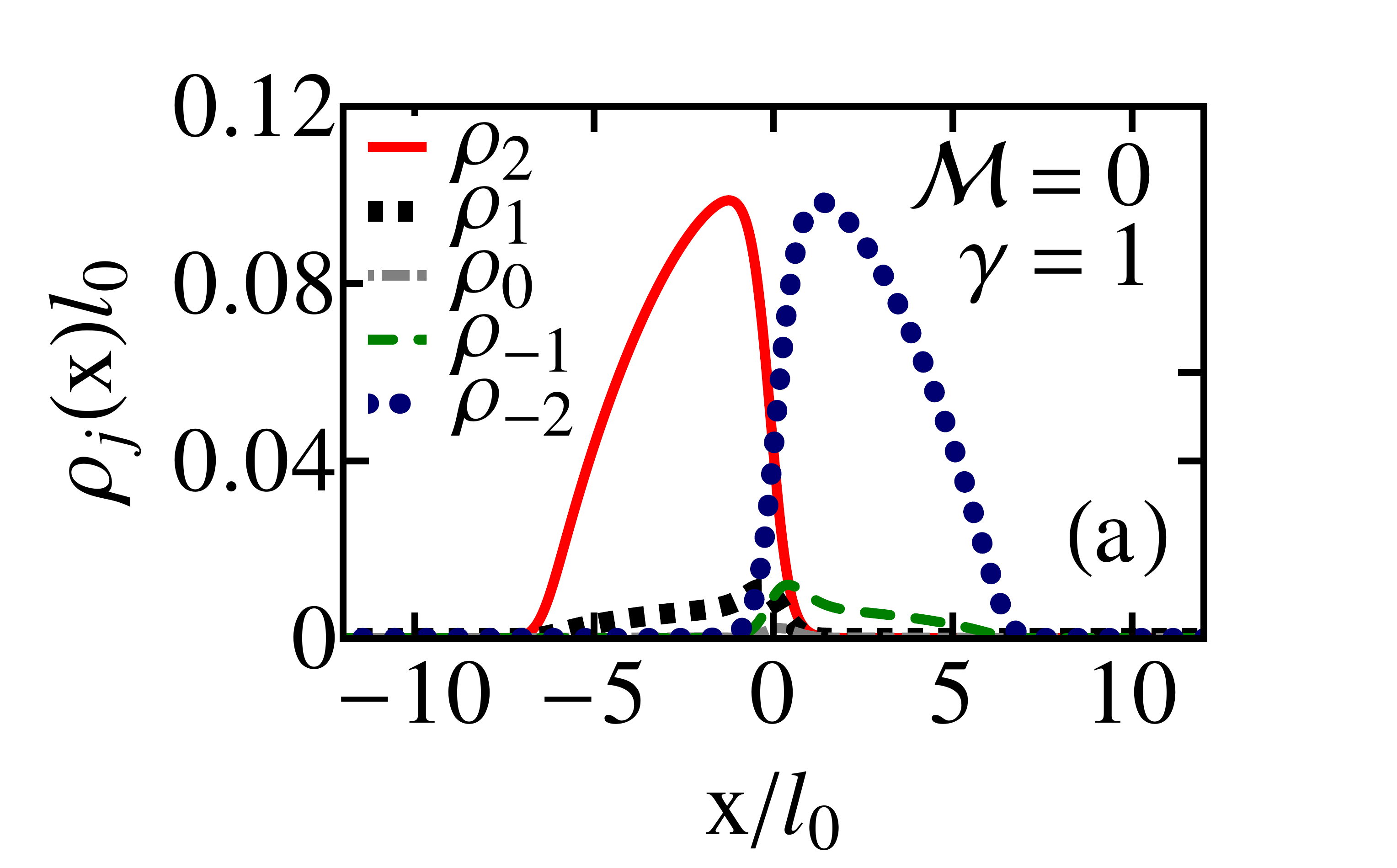}
\includegraphics[trim = 5mm 0mm 4cm 0mm, clip,width=.48\linewidth,clip]{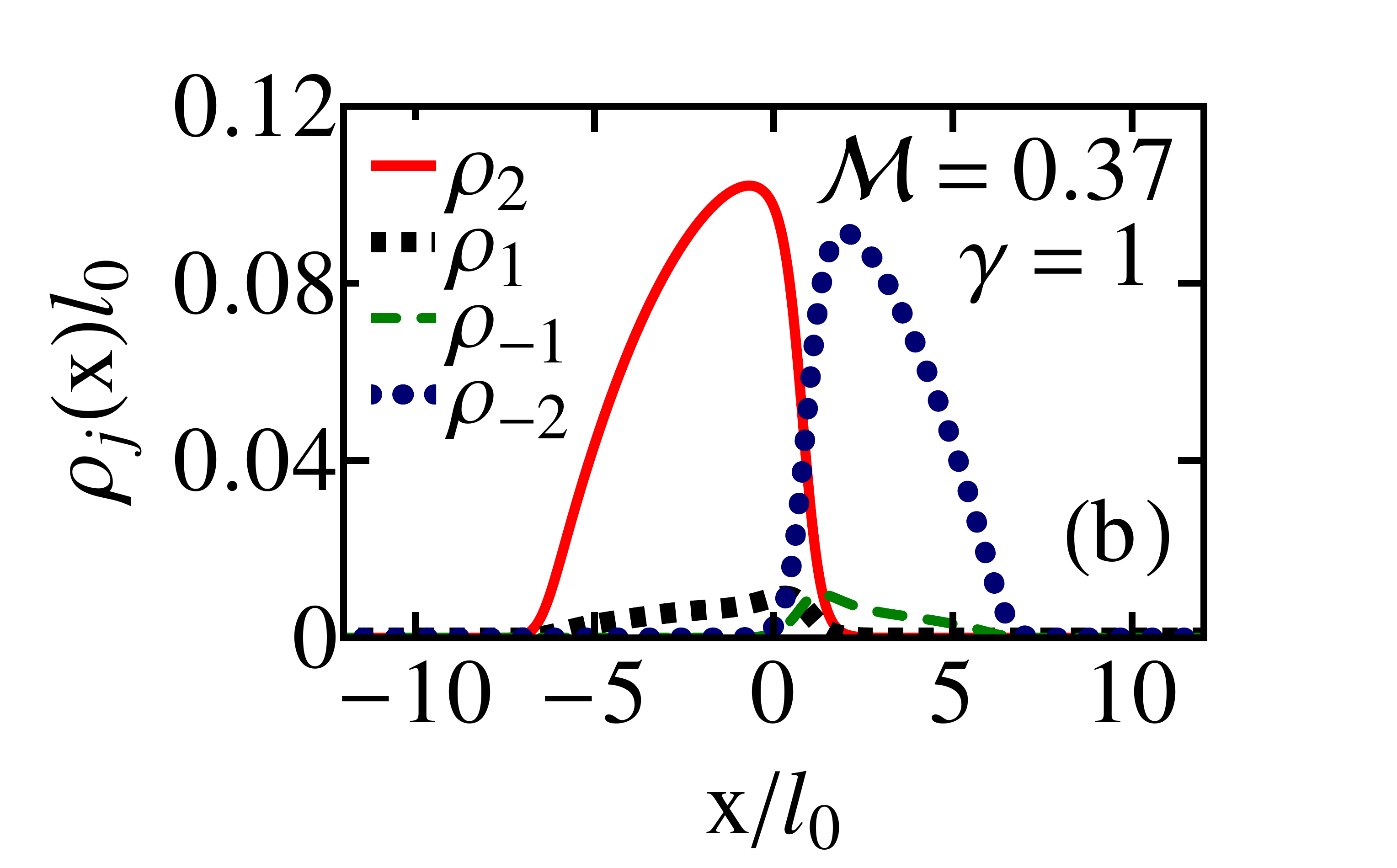}
\includegraphics[trim = 5mm 0mm 4cm 0mm, clip,width=.48\linewidth,clip]{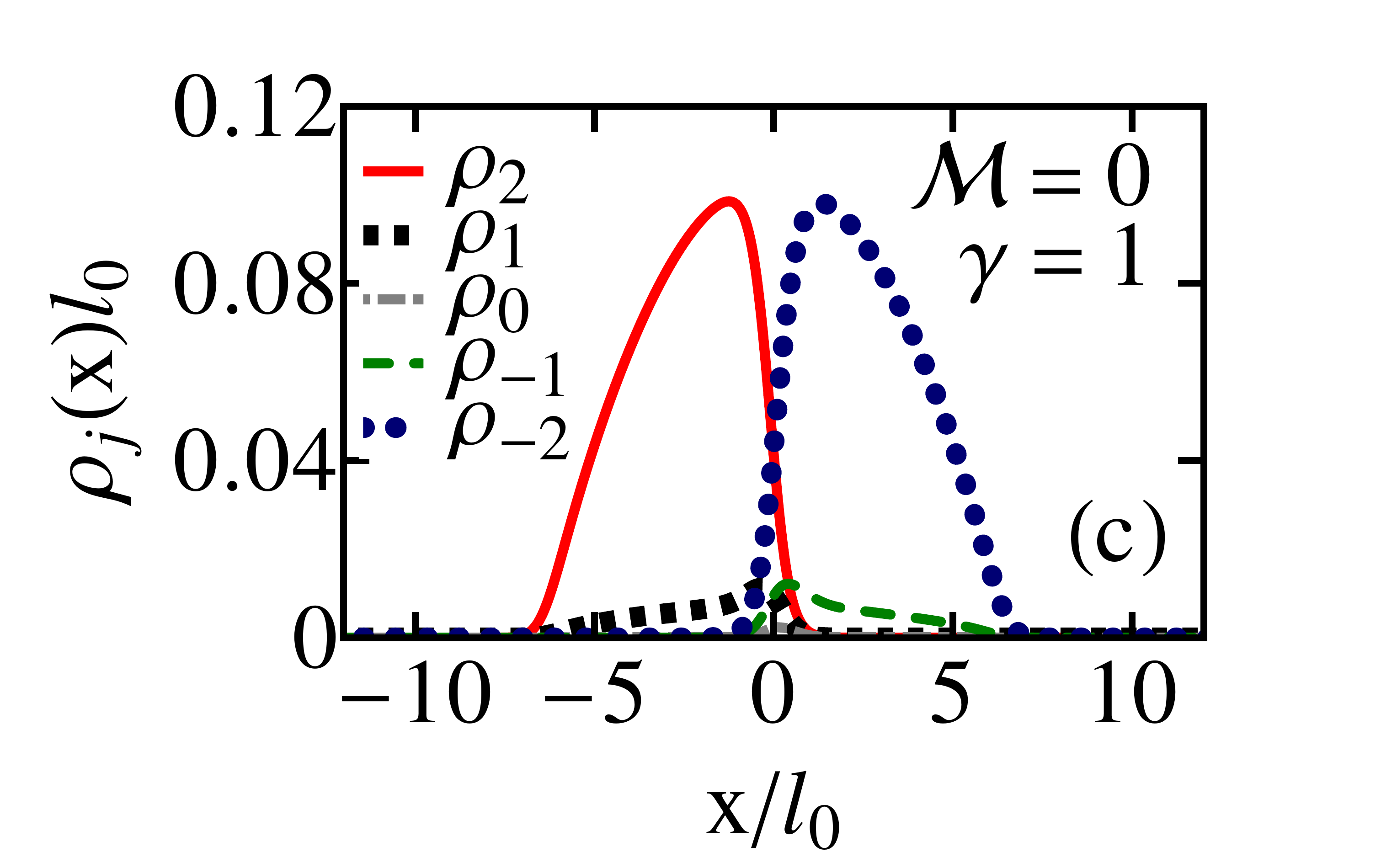}
\includegraphics[trim = 5mm 0mm 3.9cm 0mm, clip,width=.48\linewidth,clip]{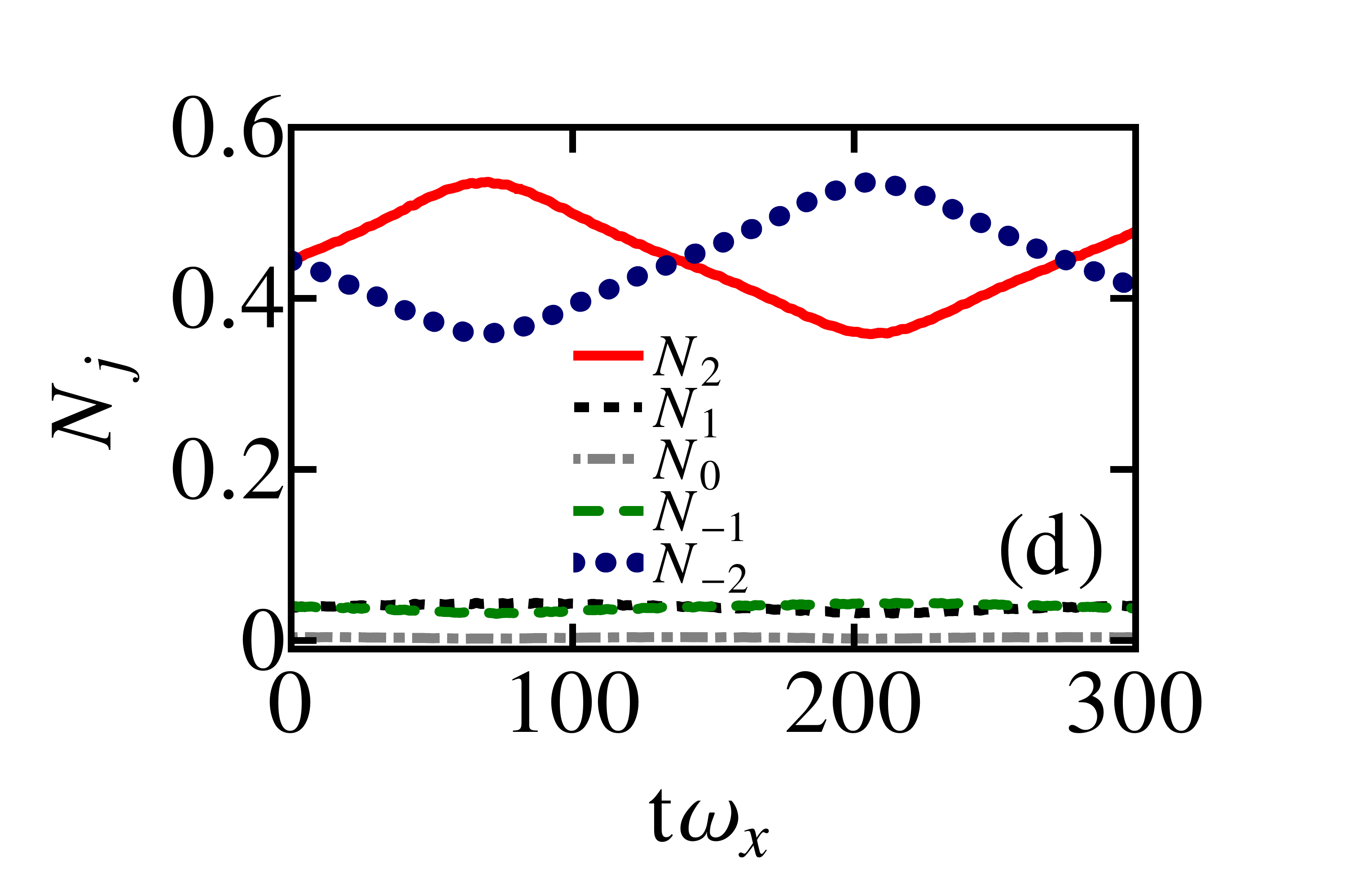}
\caption{Dynamics of a phase-separated spinor condensate in the presence of both SO 
	coupling and Rabi term ($\gamma = 1.0, \Omega = 0.5$ and a zero initial magnetization. 
(a)  The density profiles  at $t = 0$, (b) 
 the densities at $\omega_x t = 70.0$, and (c)  the densities 
 at $\omega_x t=275$, the time after which the condensate has recovered 
its initial shape after  one complete oscillation. (d) The component population $N_j =\int  \rho_j(x)dx$
versus time during a complete cycle of the periodic oscillation. }
\label{fig-8} \end{center}
\end{figure}

{\em Spin-mixing dynamics in a phase-separated spinor condensate:} In the presence
of a Rabi term ($\Omega\ne 0$) the solutions of Eqs. (\ref{gp_s1}) - (\ref{gp_s5}), in general,  
are not stationary and exhibit oscillating  spin-mixing dynamics \cite{H_Wang,Pu}. To 
study the spin-mixing dynamics in a phase-separated spinor condensate, we again consider 
$10000$ atoms of $^{23}$Na with $a_0 = 52.35a_B$, $a_2 = 45.8a_B$ $a_4 = 43.0a_B$,  as in Fig. 
\ref{fig-2},
yielding $c_0 = 201.36, c_1=-1.81, c_2 = 24.15 > 20c_1$.  We first solve 
Eqs. (\ref{gp_s1}) - (\ref{gp_s5}) using imaginary-time propagation employing $\gamma = 1, 
{\cal M }=0$ and $\Omega = 0.5$
with the 
aforementioned set of parameters. The  initial component densities 
so  obtained are  shown in 
Fig. \ref{fig-8} (a). We then evolve this solution using real-time propagation. 
The presence of the Rabi term leads to spin-mixing between the various components. 
The magnetization  ${\cal M }$ is no longer a conserved
parameter in the presence of the Rabi term as can be interpreted from Fig. \ref{fig-8} (b),
where the magnetization is $0.37$ at time $\omega_x t=70$. During the evolution, the condensate
densities execute  oscillation
periodically recovering the initial shape as shown in Fig.
 \ref{fig-8} (c) at time $\omega_xt = 275$ after one 
complete oscillation.  In Fig.  \ref{fig-8} (d) we plot the component normalizations $N_j$ versus time 
during this periodic oscillation.

\section{Summary of results}
\label{Sec-IV}

We studied the density profile of a trapped SO-coupled $f=2$ BEC for different values of the 
inter-atomic scattering lengths. Such modification of the scattering lengths can be realized in a 
laboratory using the Feshbach resonance technique \cite{fesh}.  The Hamiltonian of this problem 
preserves time-reversal symmetry but breaks parity. The wave functions of the different spin 
components are complex in general. In the anti-ferromagnetic domain only miscible density profile 
of different components are found. In this domain the wave functions preserve time-reversal symmetry. 
In the ferromagnetic domain phase-separated  density profile 
of different components are found. The underlying wave functions could be degenerate  and break time reversal  symmetry in this domain with time-reversal operator connecting two degenerate wave functions.
A class of parity-breaking states are found where the real and imaginary parts of wave functions 
exhibit opposite parities.   These conclusions were illustrated by a numerical solution of a mean-field model


\begin{acknowledgements}
This work is financed by the  Funda\c c\~ao de Amparo \`a Pesquisa do Estado de S\~ao Paulo (Brazil)
 under Contract Nos. 2013/07213-0, 2012/00451-0 and 
also   by the Conselho Nacional de Desenvolvimento Cient\'ifico e Tecnol\'ogico (Brazil).
\end{acknowledgements}


\begin{thebibliography}{99}
\bibitem{Stamper-Kurn}
 D.~M.~Stamper-Kurn, M.~R.~Andrews, A.~P.~Chikkatur, S.~Inouye, H.-J.~Miesner, 
 J.~Stenger, and W.~Ketterle, 
 Phys. Rev. Lett. {\bf 80}, 2027 (1998);
 J. Stenger, S. Inouye, D. M. Stamper-Kurn, H.-J. Miesner, A. P. Chikkatur, and 
 W. Ketterle,
 Nature {\bf 396}, 345 (1998).
\bibitem{ueda}
 Y. Kawaguchi and M. Ueda, 
 Phys. Rep. {\bf 520}, 253 (2012).
\bibitem{kurn-ueda}
 D.~M.~Stamper-Kurn and M. Ueda,
 Rev. Mod. Phys. {\bf 85}, 1191 (2013).
\bibitem{Ohmi}
 T.~Ohmi, and K.~Machida, 
 J. Phys. Soc. Japan, {\bf 67}, 1822 (1998).
\bibitem{Ho}
 T.~L.~Ho, 
 Phys. Rev. Lett. {\bf 81}, 742 (1998).
\bibitem{Koashi}
 M. Koashi and M. Ueda, 
 Phys. Rev. Lett. {\bf 84}, 1066 (2000).
\bibitem{Ciobanu}
 C. V. Ciobanu, S.-K. Yip, and T.-L. Ho, 
 Phys. Rev. A {\bf 61}, 033607 (2000).
\bibitem{sandeep}
 S. Gautam and S. K. Adhikari, \pra {\bf 90}, 043619 (2014).
\bibitem{luca}   
 L. Salasnich, A. Parola, and L. Reatto, 
 Phys. Rev. A {\bf 65}, 043614 (2002). 
\bibitem{Osterloh}
 K. Osterloh, M. Baig, L. Santos, P. Zoller, and M. Lewenstein,
 Phys. Rev. Lett. 95, 010403 (2005);
 J. Ruseckas, G. Juzeli\={u}nas, P. \"Ohberg, and M. Fleischhauer,
 Phys. Rev. Lett. {\bf 95}, 010404 (2005).
\bibitem{young}
 J. Higbie and D. M. Stamper-Kurn, 
 Phys. Rev. Lett. {\bf 88}, 090401 (2002); 
 T. L. Ho and S. Zhang, Phys. Rev. Lett. {\bf 107}, 150403 (2011); 
 Y. Deng, J. Cheng, H. Jing, C. P. Sun, and S. Yi, 
 Phys. Rev. Lett. {\bf 108}, 125301 (2012); 
 J. Radic, T. A. Sedrakyan, I. B. Spielman, and V. Galitski,
 Phys. Rev. A {\bf 84}, 063604 (2011).
\bibitem{Bychkov}
 Y.~A.~Bychkov and E.~I.~Rashba, 
 J. Phys. C {\bf 17}, 6039 (1984).
\bibitem{Dresselhaus}
 G.~Dresselhaus, 
 Phys. Rev. {\bf 100}, 580 (1955).
\bibitem{Liu}
 X.-J. Liu, M.~F.~Borunda, X.~Liu, and J.~Sinova,
 Phys. Rev. Lett. {\bf 102}, 046402 (2009).
\bibitem{Lin}
 Y.-J. Lin, K.~Jim\'enez-Garc\'ia, and I.~B.~Spielman, 
 Nature {\bf 471},  83 (2011).
\bibitem{Linx}
 V. Galitski and I. B. Spielman,
 Nature {\bf 494}, 49 (2013).
\bibitem{JY_Zhang}
 J.-Y. Zhang, S.-C. Ji, Z.~Chen, L.~Zhang, Z.-D. Du, B.~Yan, G.-S.~Pan, 
 B.~Zhao,
 Y.-J.~Deng, H.~Zhai, S.~Chen, and J.-W.~Pan,
 Phys. Rev. Lett. {\bf 109}, 115301 (2012);
 C.~Qu, C.~Hamner, M.~Gong, C.~Zhang, and P.~Engels,
 Phys. Rev. A {\bf 88}, 021604(R) (2013);
 M. Aidelsburger, M. Atala, and S. Nascimb´ene et al., 
 Phys. Rev. Lett. {\bf 107}, 255301 (2011);
 Z. Fu, P. Wang, and S. Chai, L. Huang, and J. Zhang,
 Phys. Rev. A {\bf 84}, 043609 (2011).
\bibitem{Juzeliunas}
 G.~Juzeli\={u}nas, J.~Ruseckas, and J.~Dalibard,
 Phys. Rev. A {\bf 81}, 053403 (2010);
 J. Dalibard {\em et al.}, 
 Rev. Mod. Phys. {\bf 83}, 1523 (2011).
\bibitem{lan} Z. Lan and P. \"Ohberg, \pra {\bf 89,} 023630 (2014). 
\bibitem{P_Wang}
 P.~Wang, Z.-Q.~Yu, Z.~Fu, J.~Miao, L.~Huang,
 S.~Chai, H.~Zhai, and J.~Zhang,
 Phys. Rev. Lett. {\bf 109}, 095301 (2012);
 L.~W.~Cheuk, A.~T.~Sommer, Z.~Hadzibabic, T.~Yefsah, W.~S.~Bakr, 
 and M.~W.~Zwierlein,
 Phys. Rev. Lett. {\bf 109}, 095302 (2012).
\bibitem{Wang}
 C.~Wang, C.~Gao, C-M Jian, and H.~Zhai,
 Phys. Rev. Lett. {\bf 105}, 160403 (2010); A.
 Aftalion and P. Mason, Phys. Rev. A 88, 023610 (2013);
 R. Gupta, G. S. Singh, and J. Bosse,
 Phys. Rev. A {\bf 88}, 053607 (2013);
 Q.-Q. Lu and D. E. Sheehy,
 Phys. Rev. A {\bf 88}, 043645 (2013).
\bibitem{Gopalakrishnan}
 T.~D.~Stanescu, B.~Anderson, and V.~Galitski,
 Phys. Rev. A {\bf 78}, 023616 (2008);
 S.-K. Yip, 
 Phys. Rev. A {\bf 83}, 043616 (2011);
 C.-J Wu, I.~Mondragon-Shem, and X.-F. Zhou,
 Chin. Phys. Lett. {\bf 28}, 097102 (2011);
 Q. Zhou and X. Cui,
 Phys. Rev. Lett. {\bf 110}, 140407 (2013);
 S.~Gopalakrishnan, A.~Lamacraft, and P.~M.~Goldbart,
 Phys. Rev. A {\bf 84}, 061604(R) (2011);
 H.~Hu, B.~Ramachandhran, H.~Pu, and X.-J. Liu,
 Phys. Rev. Lett. {\bf 108}, 010402 (2012);
 B.~Ramachandhran, B.~Opanchuk, X.-J.~Liu, H.~Pu, P.~D.~Drummond, and H.~Hu,
 Phys. Rev. A {\bf 85}, 023606 (2012);
 S.~Sinha, R.~Nath, and L.~Santos, 
 Phys. Rev. Lett. {\bf 107}, 270401 (2011);
 T.~Ozawa and G.~Baym, 
 Phys. Rev. A {\bf 85}, 013612 (2012);
 Y.~Zhang, L.~Mao, and C.~Zhang,
 Phys. Rev. Lett. {\bf 108}, 035302 (2012);
 K. Riedl, C. Drukier, P. Zalom, and P. Kopietz,
 Phys. Rev. A {\bf 87}, 063626 (2013);
 Y. Deng, J. Cheng, H. Jing, and S. Yi,
 Phys. Rev. Lett. {\bf 112}, 143007 (2014).
\bibitem{Ruokokoski}
 E.~Ruokokoski, J.~A.~M.~Huhtam\"{a}ki, and M.~M\"{o}tt\"{o}nen,
 Phys. Rev A {\bf 86}, 051607(R) (2012);
 S.-W.~Su, I.-K.~Liu, Y.-C.~Tsai, W.~M.~Liu, and S.-C.~Gou, 
 Phys. Rev. A {\bf 86}, 023601 (2012);
 S.-W.~Song, Y.-C. Zhang, L.~Wen, and H.~Wang,
 J. Phys. B {\bf 46}, 145304 (2013);
 S.-W. Song, Y.-C. Zhang, H. Zhao, X. Wang, and W.-M. Liu,
 Phys. Rev. A {\bf 89}, 063613 (2014).
\bibitem{Kawakami}
 T.~Kawakami, T.~Mizushima, and K.~Machida, 
 Phys. Rev. A {\bf 84}, 011607(R) (2011);
 Z.~F.~Xu, R.~L\"{u}, and L.~You, 
 Phys. Rev. A {\bf 83}, 053602 (2011);
 Z.~F.~Xu, Y.~Kawaguchi, L.~You, and M.~Ueda, 
 Phys. Rev. A {\bf 86}, 033628 (2012).
\bibitem{Zhou}
 F. Zhou,
 Phys. Rev. Lett. 87, 080401 (2001); 
 W.~Zhang, S.~Yi, and L.~You,
 New J. Phys. {\bf 5}, 77 (2003);
 K.~Murata, H.~Saito, and M.~Ueda,
 Phys. Rev. A {\bf 75}, 013607 (2007);
\bibitem{Matuszewski}
 M.~Matuszewski, T.~J.~Alexander, and Y.~S.~Kivshar,
 Phys. Rev. A 80, 023602 (2009);
 M.~Matuszewski,
 Phys. Rev. A {\bf 82}, 053630 (2010).
\bibitem{GP-Zheng}
 G.-P. Zheng, Y.-G. Tong, and F.-L. Wang,
 Phys. Rev. A {\bf 81}, 063633 (2010);
 H. Saito and M. Ueda,
 Phys. Rev. A {\bf 72}, 053628 (2005).
\bibitem{josep}
 M. A. Garcia-March, G. Mazzarella, L. Dell'Anna, 
 B. Juli\'a-D\'iaz, L. Salasnich, and A. Polls,
 Phys. Rev. A {\bf 89}, 063607 (2014).
\bibitem{Larson}
 J.~Larson, J.-P.~Martikainen, A.~Collin, and E.~Sj\"{o}qvist,
 Phys. Rev. A 82, 043620 (2010).

\bibitem{sol}
 H. Sakaguchi, Ben Li, and B. A. Malomed,
 Phys. Rev. E {\bf 89}, 032920 (2014);
 Y. Xu, Y. Zhang, and B. Wu, Phys. 
 Rev. A {\bf 87}, 013614 (2013); 
 O. Fialko, J. Brand, and U. Z¨ulicke, 
 Phys. Rev. A {\bf 85}, 051605(R) (2012);
 Y.-K. Liu and S.-J. Yang,
 Euro. Phys. Lett. {\bf 108}, 30004 (2014).
\bibitem{He}
 P.-S. He, Y.-H. Zhu, and W.-M. Liu
 Phys. Rev. A {\bf 89}, 053615 (2014).


\bibitem{Merkl}
 M. Merkl, A. Jacob, F. E. Zimmer, P. \"{O}hberg, and L. Santos,
 Phys. Rev. Lett. {\bf 104}, 073603 (2010).
\bibitem{super} 
 T. Ozawa, L. P. Pitaevskii, and S. Stringari,
 Phys. Rev. A {\bf 87}, 063610 (2013);
 D. W. Zhang, J. P. Chen, C. J. Shan, Z. D. Wang, and
 S. L. Zhu, Phys. Rev. A {\bf 88}, 013612 (2013); 
 Q. Zhu, C. Zhang and B. Wu, 
 Europhys. Lett. {\bf 100}, 50003 (2012);
 D. Toniolo and J. Linder,
 Phys. Rev. A {\bf 89}, 061605(R) (2014).


\bibitem{H_Wang}
 H.~Wang,
 J. Comput. Phys.,
 {\bf 230}, 6155 (2011); 
 {\bf 274}, 473 (2014).



\bibitem{Y_Zhang}
 Y. Li, G. I. Martone, L. P. Pitaevskii, and S. Stringari,
 Phys. Rev. Lett. {\bf 110}, 235302 (2013);
 Y.~Zhang and C.~Zhang,
 Phys. Rev. A {\bf 87}, 023611 (2013);
 L.~Salasnich and B.~A.~Malomed,
 Phys. Rev. A {\bf 87}, 063625 (2013);
D. A. Zezyulin, R. Driben, V. V Konotop, and B. A. Malomed,
 Phys. Rev. A {\bf 88}, 013607  (2013);
 Y. Cheng, G. Tang, and S. K. Adhikari,
 Phys. Rev. A {\bf 89}, 063602 (2014). 

\bibitem{Ao}
 P.~Ao and S.~T.~Chui,
 Phys. Rev. A {\bf 58}, 4836 (1998);
 P. Facchi, G. Florio, S. Pascazio, and F. V. Pepe,
 J. Phys. A: Math. Theor. {\bf 44} 505305 (2011).
\bibitem{Gautam}
 S. Gautam and D. Angom,
 J. Phys. B {\bf 44}, 025302 (2011);
 S. Gautam and D. Angom,
 J. Phys. B {\bf 43}, 095302 (2010).
\bibitem{Muruganandam}
 P.~Muruganandam and S.~K.~Adhikari,
 Comput. Phys. Commun. {\bf 180}, 1888 (2009);
 D. Vudragovic, I. Vidanovic, A. Balaz, P. Muruganandam, and S. K. Adhikari.
 Comput. Phys. Commun. {\bf 183}, 2021 (2012). 
\bibitem{Bao}
  W.~Bao and F.~Y.~Lim,
  Siam J. Sci. Comp. {\bf 30}, 1925 (2008);
  F.~Y.~Lim and W.~Bao,
  Phys. Rev. E {\bf 78}, 066704 (2008).
\bibitem{Cngf}W. Bao and Q. Du, 
  Siam J. Sci. Comp.
{\bf 25}, 1674 (2004).
\bibitem{num_recipies}
 W. H. Press. S. A Teukolsky, W. T. Vetterling, and B. P. Flannrey,
 {\em Numerical Recipies in Fortran 77 }, Cambridge University Press, 2nd Edition, (1992).
\bibitem{Esry}
 B. D. Esry and C. H. Greene,
 Phys. Rev. A {\bf 59}, 1457 (1999);
 S. T. Chui and P. Ao,
 Phys. Rev. A {\bf 59}, 1473 (1999).
\bibitem{Pu}
 H. Pu, C. K. Law, S. Raghavan, J. H. Eberly, and N. P. Bigelow,
 Phys. Rev. A {\bf 60}, 1463 (1999). 


\bibitem{fesh} 
 S. Inouye {\it et al.}, Nature (London) {\bf 392}, 151 (1998).
 





\end{thebibliography}
\end{document}